\documentclass[a4,12pt]{article}
\usepackage{jheppub}
\usepackage{caption}
\usepackage{subcaption}
\usepackage{lscape}
\usepackage{makecell}
\usepackage{amsmath,amssymb,amsfonts, graphics,graphicx,amsfonts,amscd,epsf,epsfig,dsfont,color,braket}
\usepackage[section]{placeins}

\newcommand{\nn}{\nonumber}
\newcommand{\p}{\partial}
\newcommand{\Tr}[1]{\:{\rm Tr}\,#1}

\allowdisplaybreaks[4]

\subheader{LLNL-JRNL-840804, DMUS-MP-22/15, RIKEN-iTHEMS-Report-22, YITP-22-103} 
\title{Precision test of gauge/gravity duality in D0-brane matrix model  at low temperature}
\collaboration{Monte Carlo String/M-theory Collaboration (MCSMC)}

\author[m]{Stratos Pateloudis,}
\author[a]{Georg Bergner,}
\author[t]{Masanori Hanada,}
\author[r]{Enrico Rinaldi,}
\author[m]{Andreas Sch\"{a}fer,}
\author[i]{Pavlos Vranas,}
\author[x]{Hiromasa Watanabe,}
\author[m]{Norbert Bodendorfer}

\affiliation[m]{
	University of Regensburg, Institute of Theoretical Physics,\\
	Universit\"{a}tsstrasse 31, D-93053 Regensburg, Germany}
\affiliation[a]{
	University of Jena, Institute for Theoretical Physics,\\
	Max-Wien-Platz 1, D-07743 Jena, Germany}
\affiliation[t]{
	Department of Mathematics, University of Surrey, Guildford, Surrey, GU2 7XH, United Kingdom}
\affiliation[r]{
	Physics Department, University of Michigan, Ann Arbor, MI 48109, United States\\
Theoretical Quantum Physics Laboratory, Cluster for Pioneering Research, RIKEN, Wako,
Saitama 351-0198, Japan\\
Interdisciplinary Theoretical \& Mathematical Science Program (iTHEMS), RIKEN, Wako, Saitama 351-0198, Japan\\
Center for Quantum Computing (RQC), RIKEN, Wako, Saitama 351-0198, Japan}
\affiliation[i]{
	Physical and Life Sciences Division, Lawrence Livermore National Laboratory,
	Livermore CA 94550, United States\\
Nuclear Science Division, Lawrence Berkeley National Laboratory, Berkeley, California 94720, USA
}
\affiliation[x]{
	Yukawa Institute for Theoretical Physics, Kyoto University, Kyoto 606-8502, Japan}

\abstract{We test the gauge/gravity duality between the matrix model and type IIA string theory at low temperatures with unprecedented accuracy.
	To this end, we perform lattice Monte Carlo simulations of the Berenstein-Maldacena-Nastase (BMN) matrix model, which is the one-parameter deformation of the Banks-Fischler-Shenker-Susskind (BFSS) matrix model, taking both the large $N$ and continuum limits. 
	We leverage the fact that sufficiently small flux parameters in the BMN matrix model have a negligible impact on the energy of the system while stabilizing the flat directions so that simulations at smaller $N$ than in the BFSS matrix model are possible.
	Hence, we can perform a precision measurement of the large $N$ continuum energy at the lowest temperatures to date. 
	The energy is in perfect agreement with supergravity predictions including estimations of $\alpha'$-corrections from previous simulations. 
	At the lowest temperature where we can simulate efficiently ($T=0.25\lambda^{1/3}$, where $\lambda$ is the 't Hooft coupling), the difference in energy to the pure supergravity prediction is less than $10\%$. 
	Furthermore, we can extract the coefficient of the $1/N^4$ corrections at a fixed temperature with good accuracy, which was previously unknown.}

\date{}
\begin{document}
\maketitle

\section{Introduction}\label{sec:intro}
Gauge/gravity duality was originally formulated in terms of D$p$-branes~\cite{Maldacena:1997re,Itzhaki:1998dd}. 
In the decoupling limit, the system of D$p$-branes in superstring theory admits two descriptions: weakly-coupled string theory and strongly-coupled gauge theory. 
Supergravity is a good approximation to the large-$N$ and strong-coupling limit on the gauge theory side. 
As solutions to the ten-dimensional Einstein equation, black $p$-brane geometries are known. 

Gauge/gravity duality relates the U($N$) gauge theory on the worldvolume of the D$p$-branes to superstring theory on the black $p$-branes geometry~\cite{Itzhaki:1998dd}. 
It is conjectured that the duality is valid at nonperturbative level, including the finite-$N$ and finite-coupling corrections. 
Since the strongly-coupled regime of gauge theory is dual to the weakly-coupled regime of string theory, a simultaneous study of both regions is impossible perturbatively. 
This fact motivates us to try numerical approaches on the gauge theory side to solve the theory fully non-perturbatively. 

For numerical simulations, $p=0$ is the most convenient. 
The gravity dual in the 't Hooft large-$N$ limit is conjectured to be the black zero-brane in type IIA supergravity~\cite{Itzhaki:1998dd}.\footnote{At stronger coupling region, we expect to see the M-theory~\cite{Banks:1996vh,deWit:1988wri,Itzhaki:1998dd,Bergner:2021goh}.
	In this paper, we will focus on the 't Hooft large-$N$ limit and type IIA superstring theory.
}
The dual gauge theory is matrix quantum mechanics and it can be put on a computer quite efficiently since one has to deal with just one lattice dimension.

Numerical simulation can play the important role in either falsifying or verifying the conjecture~\cite{Anagnostopoulos:2007fw, Catterall:2008yz, Hanada:2008ez,Hanada:2008gy,Hanada:2013rga,Kadoh:2015mka, Filev:2015hia,Berkowitz:2016jlq,Rinaldi:2017mjl}.
\footnote{There are attempts to test the duality at $p>0$ via lattice simulations \cite{Catterall:2010fx, Catterall:2017lub, Catterall:2020nmn}.} 
It remains an outstanding challenge to reproduce either analytically or numerically the exact gravitational results in the strong coupling limit of the gauge theory with good accuracy. 
In this paper, we elaborate on the numerical study of the D0-brane matrix model and show good agreement with superstring theory. 
We take the large-$N$ limit, where the $g_s$ corrections disappear, and study the rather strong-coupling regime where the $\alpha'$-corrections are small.  

To this aim, we shall use the Berestein-Maldacena-Nastase (BMN) model~\cite{Berenstein:2002jq} which is a massive deformation of the massless D0-brane (BFSS) matrix model~\cite{Banks:1996vh,Itzhaki:1998dd}. 
We put this matrix model in a test against the semi-analytic results obtained in Ref.~\cite{Costa:2014wya} and we reproduce the correct scaling of the internal energy of the gravitational system at low temperatures, which is the strongly-coupled regime of this model. 
Concerning numerical simulations, the usefulness of this specific model arises because the flat direction is under better control and simulations become stable at lower temperatures~\cite{Bergner:2021goh,Dhindsa:2022vch,Schaich:2022duk,Pateloudis:2022GvsU}. 
At the same time, a drawback is that the dual geometry is not known analytically.
Still, significant steps were taken numerically~\cite{Costa:2014wya}. 

This paper is organized as follows: in the next section, we are presenting in a qualitative and intuitive language the main result of the paper.
We continue in Sec.~\ref{sec:theory}, by discussing a detailed theoretical analysis of both the quantum mechanical matrix models and their gravitational duals, as well as their thermodynamics. 
The lattice setup for the simulations is introduced in Sec.~\ref{sec:lattice-regularization}, while in Sec.~\ref{sec:numerical_analysis} we switch to the numerical analysis presenting the extrapolation ans{\"a}tze for performing a large $N$ and continuum analysis. 
We show that this conforms to the theoretical results while presenting a precision measurement in Sec.~\ref{sec:PrecisionMeasurement}. 
Sec.~\ref{sec:conclusion} is devoted for conclusion and discussion.

\section{D0-brane matrix models at low temperature} \label{sec:MainResult}
In this section, we give an overview of our main results.
Technical details will be explained in later sections. 
The aim of the study is a quantitative comparison between the energy of D0-brane matrix models and that of the black zero-brane (Sec.\ref{sec:theory}).
Analytic results emanating from the supergravity geometry of the black zero-brane concern the large-$N$ and low-energy limits where quantum and $\alpha'$-corrections (i.e., finite-coupling corrections, or equivalently, finite-temperature corrections, in the matrix model side) are absent
\footnote{We may note though that there are one loop corrections estimated in \cite{HyakutakeQuantumNear}.}.
A first step to estimate higher-order $\alpha'$-corrections via the simulation of the matrix model was taken in Ref.~\cite{Hanada:2008ez} and the most precise estimate so far was given in Ref.~\cite{Berkowitz:2016jlq}. 
From the gravitational point of view, we are currently \textit{agnostic} for precise coefficients of $\alpha'$-corrections. 
At the same time, it is challenging to also obtain the $g_s$ corrections (i.e., the $1/N$ corrections) beyond one loop from the gravity analyses. 

To this end, we use simulations for the D0-brane matrix models and measure certain observables such as the energy and the Polyakov loop.
We simulate in the canonical ensemble such that the energy is obtained at a fixed temperature as $E=E(T)$. For a precise definition of observables, see Sec.~\ref{sec:observables}.   
In the large-$N$ limit and at low temperatures, both $g_s$ and $\alpha'$-corrections are small, and supergravity should be the precise dual description. 
Deviations from this limit provide us with the information of the $\alpha'$ and $g_s$ corrections. 
By fitting the simulation results of the matrix model, we can estimate those corrections with good precision. 

In the past, the biggest obstacle for the simulation at low temperatures was the instability associated with the flat directions (Sec.~\ref{sec:flat-direction}). 
To tame the flat direction in a cost-effective manner, we will simulate the BMN matrix model~\cite{Berenstein:2002jq} (Sec.~\ref{sec:theory}). 
This matrix model adds a deformation parameter, $\mu$, which can be considered as the mass of the bosons and fermions of the model, and which reduces the instability due to the flat directions. 
The un-deformed theory is called the BFSS matrix model. 
For sufficiently small $\mu$, the energy does not change much from the value at $\mu=0$, while the flat directions are under control. 

\begin{figure}
	\centering 
	\includegraphics[scale=0.47]{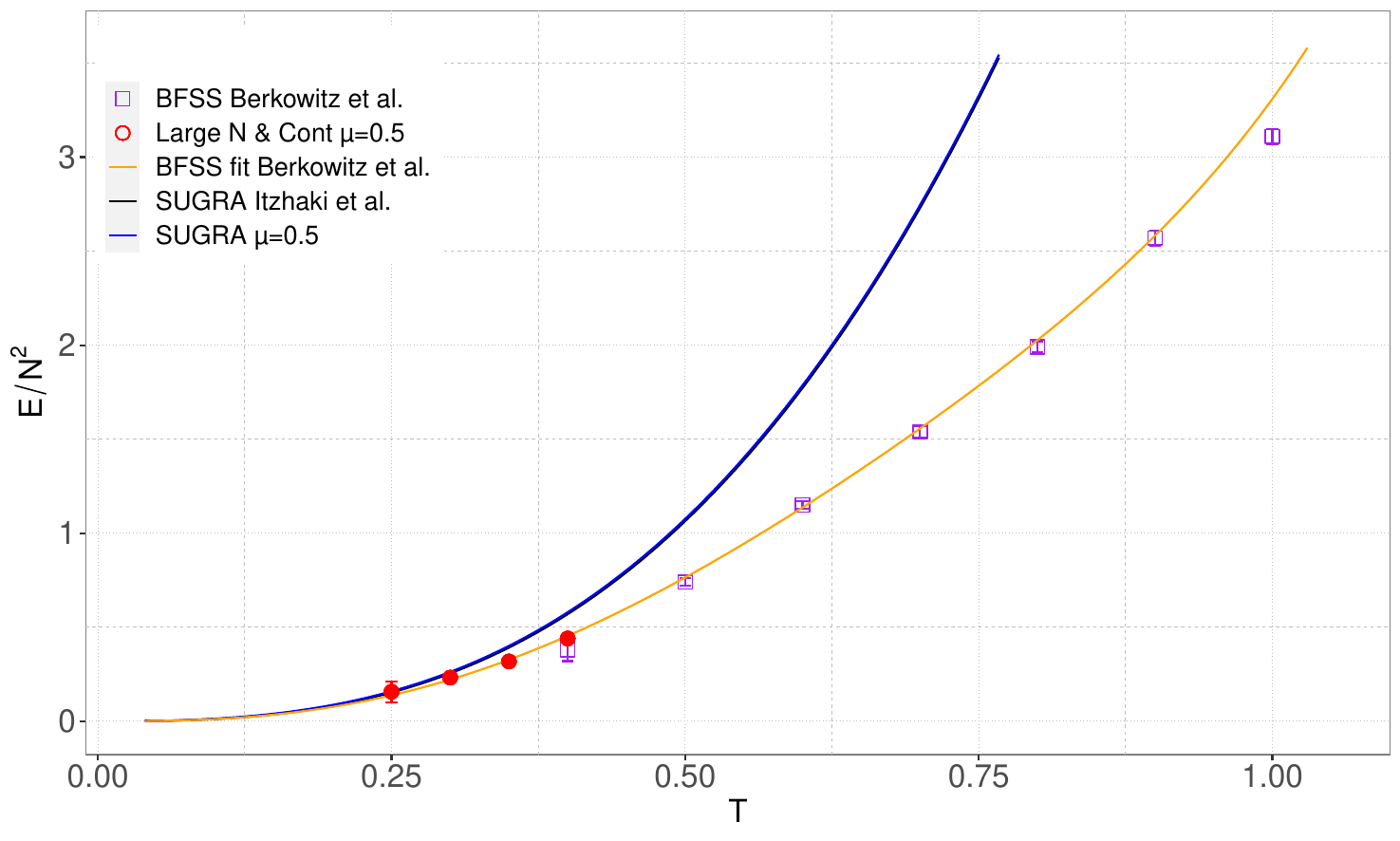}
	\caption{In addition to the values obtained in this work (red points; see Sec.~\ref{sec:numerical_analysis} for details), we show the dual supergravity prediction (black line)~\cite{Itzhaki:1998dd}, the values obtained from the BFSS matrix model in the past (purple points)~\cite{Berkowitz:2016jlq}, and the fit of the BFSS results that takes into account the $\alpha'$-corrections (orange line)~\cite{Berkowitz:2016jlq}. 
	}
	\label{fig:BFSS_vs_BMN_full} 
\end{figure}

\begin{figure}
	\centering 
	\includegraphics[scale=0.47]{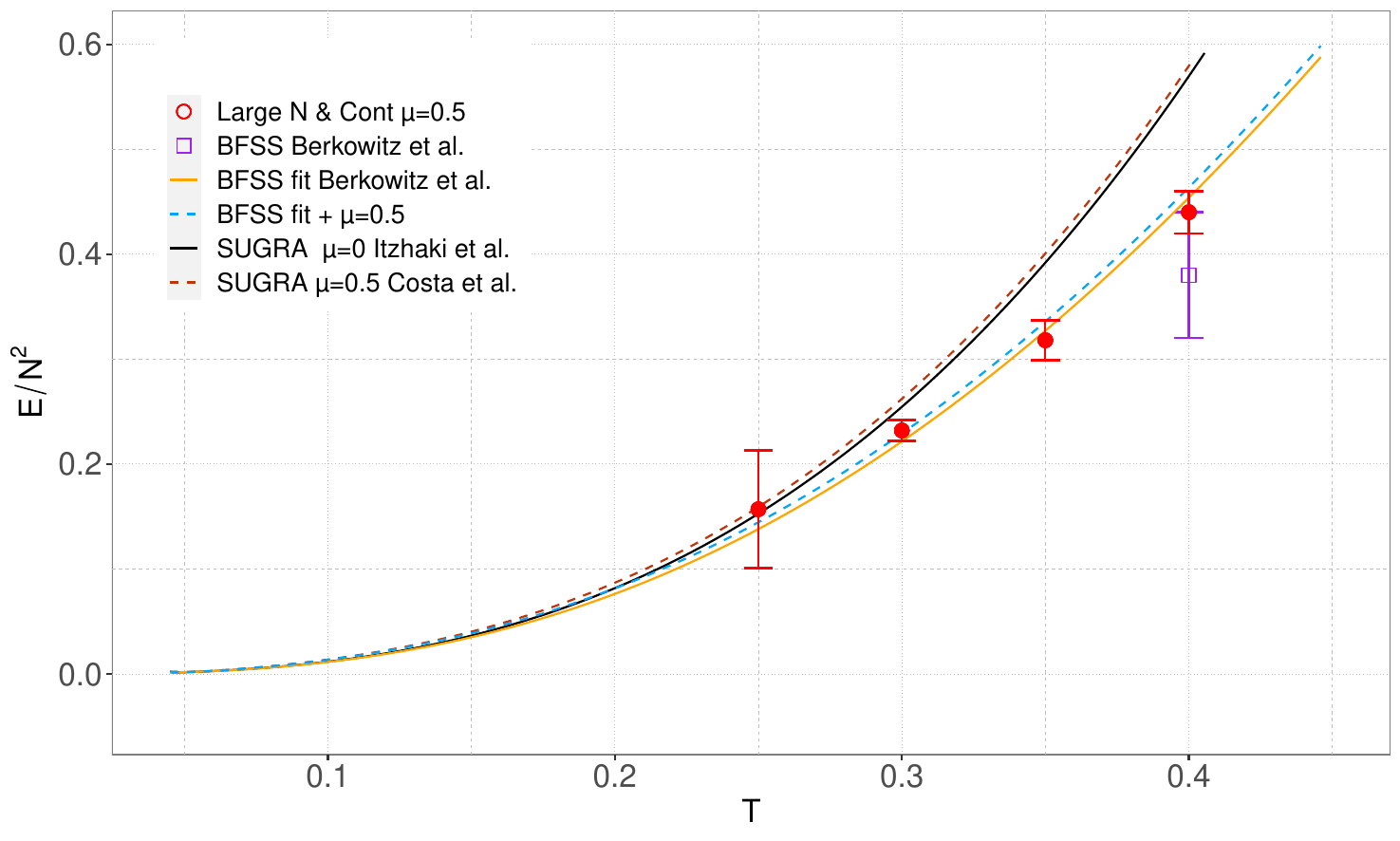}
	\caption{Zoomed-in view of the low-temperature region. The dashed lines include finite $\mu$ corrections. The blue and orange lines are based on the ansatz~\eqref{eq:BMN_Energy_expansion_large_N} including $\alpha'$-corrections from \cite{Berkowitz:2016jlq}. }
	\label{fig:BFSS_vs_BMN_full_zoom_in} 
\end{figure}

In Fig.~\ref{fig:BFSS_vs_BMN_full}, we show the energy as a function of temperature. 
In addition to the values obtained in this work (red points), we show the dual supergravity prediction (black line), the values obtained from the BFSS matrix model in the past (purple points)~\cite{Berkowitz:2016jlq}, and the fit of the BFSS results takes into account the $\alpha'$-corrections (orange line). 
We could study the low-temperature region that was not studied in the past and see the convergence to the supergravity prediction. 
A zoomed-in view of the low-temperature region is shown in Fig.~\ref{fig:BFSS_vs_BMN_full_zoom_in}.

\section{Theoretical analysis}\label{sec:theory}
In this section, we present the theoretical background one needs to understand the numerical results. 
We start by defining the matrix models and then provide its gravity dual. 
Finally, we perform thermodynamic analyses based on gravity dual and previous Monte Carlo studies so that the new results provided in later sections can be understood precisely.  
\subsection{The matrix model}
The D0-brane matrix models are defined in (0+1)-dimensions, with the dimension assigned to time $t$. 
Having matrix-valued variables gives rise to a quantum mechanical system containing matrices on the worldline.

We will be considering the plane-wave deformed theory, which is called the BMN matrix model~\cite{Berenstein:2002jq}.
The action is given by 
\begin{align}\label{eq:BMN_action}
S_{\rm BMN}=S_{\rm BFSS}+\Delta S. 
\end{align}

$S_{\rm BFSS}$ is the action of the BFSS model \cite{Banks:1996vh}: 
\begin{align}
S_{\rm BFSS}=
\frac{N}{\lambda}\int_0^\beta dt\ {\rm Tr}\left\{ \frac{1}{2}(D_t X_M)^2 -\frac{1}{4}[X_M,X_N]^2 +\frac{1}{2}\bar{\psi}^\alpha\gamma^{10} D_t\psi_\alpha -\bar{\psi}^\alpha\gamma^M_{\alpha\beta}[X_M,\psi^\beta]
\right\}. 
\end{align}

The BMN model differs from the BFSS model by the deformation terms\footnote{Our normalization for mass is different from Refs.~\cite{Berenstein:2002jq,Costa:2014wya} by a factor 3.} 
\begin{align}
\Delta S\label{eq:BMN_def}=
\frac{N}{\lambda}\int_0^\beta dt\ {\rm Tr}\Big\{
\frac{\mu^2}{2}\sum_{i=1}^3X_i^2
+\frac{\mu^2}{8}\sum_{a=4}^9X_a^2 +
i\sum_{i,j,k=1}^3\mu\epsilon^{ijk}X_iX_jX_k+\frac{3i\mu}{4}\bar{\psi}^\alpha\gamma_{123\alpha}{}^\beta\psi_\beta
\Big\} .
\end{align}

The model consists of nine $N \times N$ bosonic hermitian matrices $X_M$ ($M=1,\cdots,9$), sixteen fermionic matrices $\psi_\alpha$ ($\alpha=1,\cdots,16$) and the gauge field $A_t$. 
The $16 \times 16$ matrices $\gamma_{\alpha\beta}^M (M=1,\cdots, 9)$ are the left-handed parts of the gamma matrices in $(9+1)$-dimensions. 
$\epsilon^{ijk}$ is the structure constant of SU(2), which is totally antisymmetric, and $\epsilon^{123}=+1$. 
This theory arises as a dimensional reduction of $(9+1)$-dimensional super Yang-Mills theory with $\mathcal{N}=1$ supersymmetry or $(3+1)$-dimensional maximal super Yang-Mills theory with $\mathcal{N}=4$ supersymmetries to $(0+1)$ dimension.

Both $X_M$ and $\psi_\alpha$ are in the adjoint representation of the U($N$) gauge group, and the covariant derivative $D_t$ acts on them as $D_tX_M = \partial_t X_M -i[A_t,X_M]$ and $D_t\psi_\alpha = \partial_t\psi_\alpha -i[A_t,\psi_\alpha]$. 
The equation of motion for the gauge field $A_t$ gives rise to the Gauss constraint
\begin{align}\label{singletconstraint}
\mathcal{G}:=\frac{i N}{2\lambda} (2[\dot{X}_M, X_M]+[\bar{\psi}_\alpha,\psi_\alpha])=0\quad,\quad \dot{X}\equiv \p_t X,
\end{align}
in the $A_t=0$ gauge. 

Note that this lattice action breaks supersymmetry. Still, due to the special property in the $(0+1)$ dimension, supersymmetric continuum limit is realized. This property has been used since Refs.~\cite{Hanada:2007ti,Catterall:2007fp}. The argument for the lattice regularization was given in Ref.~\cite{Catterall:2007fp}. 

This matrix model is put on a Euclidean circle with the circumference $\beta$. 
For bosonic fields ($X$) and fermionic fields ($\psi$), we take the boundary condition to be periodic and antiperiodic, respectively.
Then, $\beta$ is the inverse of the temperature, $\beta=1/T$. 
The canonical partition function at finite temperature is defined as 
\begin{align}\label{eq:Z_gauged_path_integral}
Z_{\rm BMN}=\int[\mathcal{D}A_t][\mathcal{D}X_M][\mathcal{D}\psi_\alpha]e^{-S_{\rm BMN}[X,A_t,\psi]}.
\end{align}

The extra terms appearing in the action \eqref{eq:BMN_def} are mass terms for bosons, fermions, and interaction terms. 
These terms break the original rotational symmetry of the action according to $SO(9) \to SO(3)\times SO(6)$, since in this case $i=1,2,3$ and $a=1,\cdots,6$.

The matrix $\gamma_{123}$ appearing in the fermionic mass term of \eqref{eq:BMN_def} is chosen to be\footnote{This is the $16\times 16$ representation in ten dimensions. 
	In general, the $\gamma^I, I=1,\cdots,10$ matrices are $16\times16$ sub-matrices of the $32\times 32$ ten-dimensional Gamma matrices $\Gamma^I$.} 
\begin{align}
\gamma_{123}=\begin{pmatrix}
-i\mathbf{1}_2\otimes\mathbf{1}_4&0\\
0&i\mathbf{1}_2\otimes\mathbf{1}_4
\end{pmatrix},
\end{align}
which further simplifies the mass term to
\begin{align}
\frac{3i\mu}{4}\bar{\psi}^\alpha\gamma_{123}\psi_\alpha=\frac{3\mu}{2}\bar{\psi}^\alpha\psi_\alpha.
\end{align}
In addition, the deformation terms result in a new class of vacua labeled by representations of the $SU(2)$ group. 
In other words, one can write the deformed bosonic part containing only the index $i=1,2,3$ as 
\footnote{For a more comprehensive analysis we refer to \cite{Bergner:2021goh} and \cite{Pateloudis:2022GvsU} for this potential and more discussions on the stability obtained in the simulations.}
\begin{align}
V_{\rm SO(3)}=-\frac{1}{4}\Tr\left(\mu\epsilon^{ijk}X_k+i[X_i,X_j]\right)^2.
\end{align}

From this, it is clear that the potential is minimized for $[X_i,X_j]=i\mu\epsilon^{ijk}X_k$. 
Therefore, matrices that minimize the whole BMN potential in addition to the trivial ones (i.e, $X_i=0=X_a=\psi_\alpha$) can be written in the form 
\begin{align}\label{newvacua}
\psi_\alpha=0,\qquad X_a=0\ \ {\rm for}\ \ a = 4,\cdots, 9, \qquad X_i=\mu J_i\ \ {\rm for}\ \  i=1,2,3,
\end{align}  
where $J_i$ are the generators of ${\rm SU}(2)$. 

In the limit $\mu\to 0$, the deformation terms vanish and one expects the above model to converge to the BFSS model. 
This, however, assumes that there is no phase transition between the models, and indeed evidence until now supports this assumption~\cite{Costa:2014wya, Bergner:2021goh, Pateloudis:2022GvsU}. 
Note also that the singlet constraint \eqref{singletconstraint} is not affected by the deformation. 

We can construct several effective, dimensionless coupling constants that control different regimes of the model. 
For the BMN model, we have
\begin{align}\label{eq:mu_dimensionless_coupling}
g_{\rm eff}^{(\mu)}:=\frac{\lambda}{\mu^3},~~[\lambda]=({\rm energy})^3,~~[\mu]=({\rm energy})^1, 
\end{align}
and 
\begin{align}\label{eq:r_dimensionless_coupling}
g_{\rm eff}^{(r)}:=\frac{\lambda}{r^3},~~[\lambda]=({\rm energy})^3,~~[r]=({\rm energy})^1. 
\end{align}
In the latter, $r$ is the radial coordinate constructed from the nine spatial dimensions corresponding to nine scalar fields. 
(See Ref.~\cite{Hanada:2021ipb} for the precise construction.) 
This coupling has to be large for the supergravity description \eqref{eq:BFSSdual}, and hence it should respect the bound $r^3\lesssim \lambda$ as we shall discuss later on. 

In this paper, we study thermodynamics in the canonical ensemble.\footnote{See Ref.~\cite{Bergner:2021goh} for the thermodynamics in the microcanonical ensemble.} 
The energy is obtained as a function of temperature $T$, and we obtain another dimensionless effective coupling, 
\begin{align}\label{eq:T_dimensionless_coupling}
g_{\rm eff}^{(T)}:=\frac{\lambda}{T^3},~~[\lambda]=({\rm energy})^3,~~[T]=({\rm energy})^1. 
\end{align}
The phase diagram of the BMN matrix model in terms of $T$ and $\mu$ has been studied on the gravity side~\cite{Costa:2014wya} and the gauge theory side~\cite{Bergner:2021goh}. 
Supergravity can provide us with a good approximation to thermodynamic features of the matrix model when both $\mu$ and $T$ are small. 
\subsection{Gravity dual and thermodynamic analysis}

\subsubsection{Dual gravity analysis for the BFSS matrix model}
The gravity dual of the BFSS matrix model ($\mu=0$) at strong coupling is conjectured to be the black zero-brane in type IIA supergravity formed by $N$ D0-branes~\cite{Itzhaki:1998dd}. 
The geometry in the string frame is given as 
\begin{align}\nn
\frac{ds^2}{\alpha'}&=-H(r)^{-1/2}f(r)dt^2+H(r)^{1/2}\left(\frac{dr^2}{f(r)}+r^2d\Omega_8^2\right)\ ,\\
\nn H(r)&= \frac{240\pi^5 \lambda}{r^7}\ ,~~\lambda=g_{YM}^2N\ ,\\
\nn e^\phi&=\frac{(2\pi)^2}{240\pi^5}\frac{1}{N}\left(\frac{240\pi^5 \lambda}{r^3}\right)^{\frac{7}{4}}\ ,\\
\label{eq:BFSSdual} f(r)&=1-\left(\frac{r_0}{r}\right)^7\ .
\end{align}
The location of the horizon $r_0$ is expressed by the Hawking temperature $T$ as 
\begin{align}\label{eq:temperature}
T=\frac{7}{4\pi\sqrt{240\pi^5 \lambda}}r_0^{\frac{5}{2}}\ .
\end{align}
Equivalently, 
\begin{align}\label{eq:effective_temperature}
\left(g_{\rm eff}^{(T)}\right)^{-1/3}=\frac{7}{4\pi\sqrt{240\pi^5}}\left(g_{\rm eff}^{(r_0)}\right)^{-5/6}\ .
\end{align}

For the Bekenstein-Hawking formula to be valid, stringy corrections must be small at the horizon. 
In the 't Hooft large-$N$ limit, $g_{\rm eff}^{(T)}$ and $g_{\rm eff}^{(r_0)}$ is fixed. 
The string coupling $e^\phi$ vanishes at fixed $r$, including the horizon, $r=r_0$. 
In order for the $\alpha'$-correction to be small, $g_{\rm eff}^{(r_0)}$ must be large. 
Equivalently, $g_{\rm eff}^{(T)}$ must be large, i.e., the temperature must be sufficiently low. 

On the gauge theory side, this temperature corresponds to the circumference of the Euclidean circle on which we put our matrix, namely $\beta=\frac{1}{T}$. 
Knowing the temperature one can pursue a thermodynamic analysis and compare it with the relevant quantities of the matrix model~\cite{Itzhaki:1998dd, KlebanovEntropyOfNear, HyakutakeQuantumNear, HyakutakeQuantumMwave}. 
In particular, strictly in the supergravity limit, the entropy $S$ is given by 
\begin{align}\label{eq:entropy}
S=\frac{\mathcal{A}}{4G_N}\Big|_{r=r_0}=\frac{\Omega_8r^8 H(r)^{-\frac{1}{2}}}{4 G_N}\Big|_{r=r_0}=11.52N^2\lambda^{-\frac{3}{5}}T^{\frac{9}{5}}. 
\end{align}
Here, $\mathcal{A}$ is the area of the horizon, and $\Omega_8=\frac{2\pi^{\frac{9}{2}}}{\Gamma\left(\frac{9}{2}\right)}$ is the area of the unit eight-sphere. 
In addition, we have used the conventions $16\pi G_N=(2\pi)^7\alpha'^4 g_s^2$, with $G_N$ being the ten-dimensional Newton constant and $g_s=4\pi^2\alpha'^{3/2}\lambda/N$ the string coupling.
From the entropy $S$, by using  $dE=TdS$ the internal energy $E$ is obtained 
\begin{align}\label{eq:BFSS_energy}
\frac{E}{N^2}=7.41\lambda^{-\frac{3}{5}}T^{\frac{14}{5}}. 
\end{align}
The free energy $F$ is
\begin{align}\label{eq:BFSS_free_energy}
F=E-TS=-4.11 N^2\lambda^{-\frac{3}{5}}T^{\frac{14}{5}}. 
\end{align}

Comparison between the BFSS matrix model and the black zero-brane was explored numerically using Monte-Carlo simulations for the internal energy of the theory accessing in this way the correspondence in a non-perturbative fashion; see Refs.~\cite{Anagnostopoulos:2007fw,Catterall:2008yz} for the first simulations.
In Ref.~\cite{Berkowitz:2016jlq} the gauge/gravity duality was put to a precision test in the large-$N$ and continuum limits at $T\ge 0.4\lambda^{1/3}$. 
In particular, the corrections to \eqref{eq:BFSS_energy} were considered. Both $g_s$-correction (finite-$N$ correction) and $\alpha'$-correction (finite-$T$ correction) were studied.  
The expansion concerning $T$ and $\frac{1}{N}$ is given as\footnote{To keep things simple we will be suppressing $\lambda$'s appearing in the equations from now on, while to restore units we can always multiply with appropriate powers of $\lambda$ since $[\lambda]=(\rm energy)^3$.}
\begin{align}\label{eq:BFSS_Energy_expansion}
\frac{E}{N^2}=\frac{\left(a_0 T^{\frac{14}{5}}+a_1T^{\frac{23}{5}}+a_2 T^{\frac{29}{5}}+\cdots\right)}{N^0}+\frac{\left(b_0 T^{\frac{2}{5}}+b_1 T^{\frac{11}{5}}+\cdots\right)}{N^2}+\mathcal{O}(N^{-4})\ .
\end{align}
Only $a_0=7.41$ and $b_0=-5.77$ are known analytically. The former is obtained by using supergravity~\cite{Itzhaki:1998dd}.
The latter follows from quartic curvature corrections to the eleven-dimensional supergravity \cite{HyakutakeQuantumNear} which corresponds to one-loop correction to the effective type IIA supergravity theory.
\subsubsection{Flat directions}\label{sec:flat-direction}
A technical obstacle in the past studies of the BFSS matrix model was the instability associated with the flat direction, i.e., eigenvalues of matrices can roll to infinity.\footnote{
	Strictly speaking, the partition function of the BFSS matrix model is well-defined only when the flat direction is removed. One can take the large-$N$ limit with an explicit IR cutoff, for example by adding small but nonzero value of $\mu$, and then remove the cutoff. See e.g.~Ref.~\cite{Catterall:2009xn} regarding this issue. 
} The black zero-brane solution corresponds to the bound state of eigenvalues. At finite temperatures, such a bound state can be stably simulated only when $N$ is sufficiently large.\footnote{
	Suppose that the $(N,N)$-component escaped to infinity. For this to happen, off-diagonal entries ($(N,i)$-elements and $(i,N)$-elements, where $i=1,\cdots,N-1$) must become zero, i.e., $O(N)$-number of degrees of freedom must decouple from the dynamics. Such a process enables entropic suppression which scales like $e^{-N}$. Remarkably, such a decay of the bound state is associated with the negative specific heat, similarly to the evaporation of the Schwarzschild black hole~\cite{Berkowitz:2016znt}.
} The instability increases as the temperature is lowered, and then larger $N$ is needed for a stable simulation. But larger $N$ means a larger simulation cost. 

In this paper, we wish to perform a similar test going to even lower temperatures. To this end, we will use the BMN matrix model \cite{Berenstein:2002jq} and exploit the fact that it behaves more stably even at lower temperatures because the flat direction is lifted.\footnote{
	Note that the problem associated with the flat direction is not completely resolved, because eigenvalues can reach very far when $\mu$ is small. Still, the bound state becomes much more stable.} 
The compromise is that we have to use a complicated geometry on the gravity side since the effects of the deformation terms $\mu$ are not well under control analytically. The finite-$\mu$ effects on the phase structure were studied numerically on the gravity side~\cite{Costa:2014wya} and the matrix model side~\cite{Bergner:2021goh}, and a reasonably good agreement was observed. In this work, we are further comparing the energy at lower temperatures. 

\subsubsection{Dual gravity analysis for the BMN matrix model}
\begin{figure}[ht!]
	\centering 
	\includegraphics[scale=0.8]{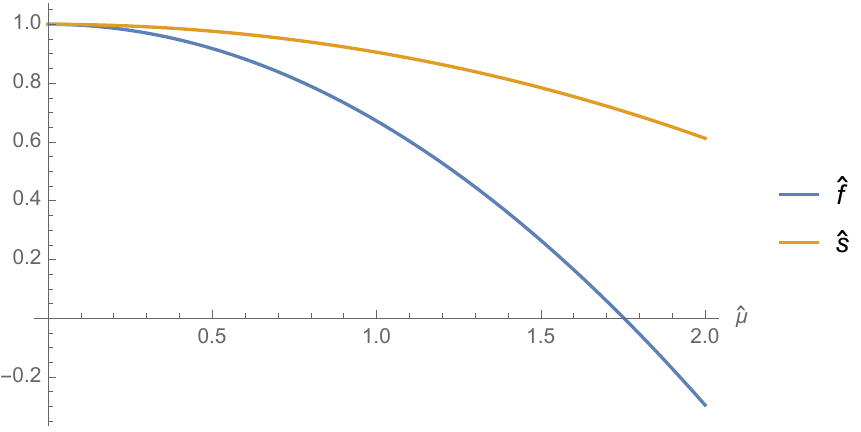}
	\caption{$f(\hat\mu)$ and $s(\hat\mu)$ from~\cite{Costa:2014wya}.}
	\label{fig:hat_fmu_smu}
\end{figure}

In this paper, we study the deconfined phase of the BFSS/BMN model in the large-$N$ limit that is dual to the black hole geometry. According to the gravity analysis~\cite{Costa:2014wya}, we should see the deconfined phase at
\begin{align}\label{eq:TcLargeN}
T\gtrsim 0.318\mu.
\end{align} 
To calculate the energy $E$ as a function of $T$ and $\mu$, we can use the free energy $F$ and entropy $S$ calculated in Ref.~\cite{Costa:2014wya} that take the following form:
\begin{align}\label{eq:BMN_thermodynamic_quantities}
F(T,\mu)\equiv f(\hat{\mu})\cdot F(T,\mu=0)\ ,
\quad 
S(T,\mu)\equiv s(\hat{\mu})\cdot S(T,\mu=0)\ ,
\quad 
\hat\mu\equiv \frac{7\mu}{4\pi T}\ .
\end{align}
The functions $f(\hat\mu)$ and $s(\hat\mu)$ capture the finite-$\mu$ corrections to the BFSS limit ($\mu=0$) in the supergravity limit. Even though they are not known in a closed form, they can be expanded by using $\hat\mu$~\cite{Costa:2014wya} as
\begin{align} \label{eq:f_function}
f(\hat\mu)=&\sum_{n=0}^\infty\frac{14s_n}{14-5n}\hat\mu^n\ ,\\
\label{eq:s_function}s(\hat\mu)=&\sum_{n=0}^\infty s_n \hat{\mu}^n\ .
\end{align} 
The coefficients $s_n$ were determined numerically to a few orders.\footnote{We would like to thank Jorge Santos for sharing some of these data with us.}
The functions $f(\hat\mu)$ and $s(\hat\mu)$ are plotted in Fig.~\ref{fig:hat_fmu_smu}. 
As we can see from \eqref{eq:BMN_thermodynamic_quantities}, the sign change of $f(\hat\mu)$ has the interpretation that at this point there is a phase transition, specifically the confinement/deconfinement transition. 

\begin{figure}[ht!]
	\centering 
	\includegraphics[scale=0.5]{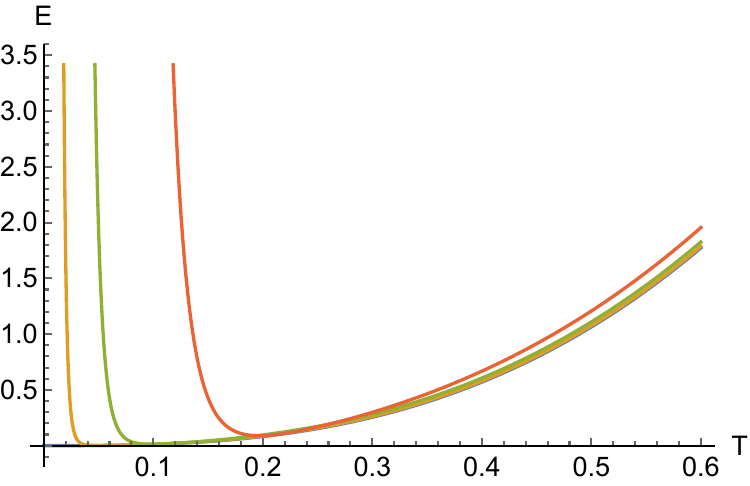}
	\includegraphics[scale=0.5]{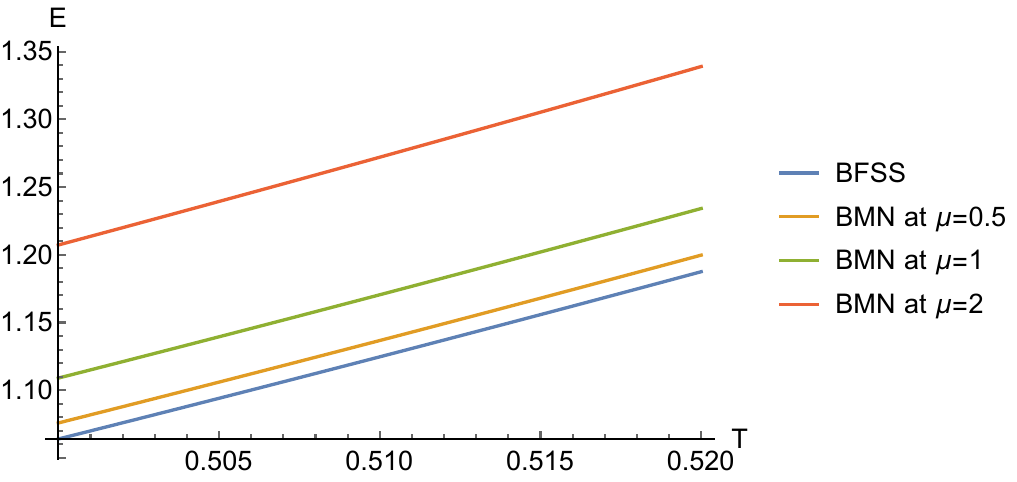}
	\caption{[Left]: The energies in the supergravity limit at fixed values of $\mu$ as functions of temperature. 
		[Right]: The same comparison zoomed in the region $0.5\le T\le 0.52$. The blowing-up behavior in the left panel is due to the truncation of the expansions \eqref{eq:f_function} and \eqref{eq:s_function}. Because the truncation of higher powers of $\hat{\mu}=\frac{7\mu}{4\pi T}$ is not valid at fixed $\mu$ and $T\to 0$, this blowing-up behavior is unphysical.}
	\label{fig:energies_comparison}
\end{figure}

\noindent
Now, we can substitute all these data to the equation of state $E=F+TS$ and get the energy in the supergravity limit, 
\begin{align}\label{eq:BMN_sugra_energy}
\frac{E(T,\mu)}{N^2}
=
-4.11 \lambda^{-\frac{3}{5}}T^{\frac{14}{5}}f(\hat\mu)+11.52\lambda^{-\frac{3}{5}}T^{\frac{14}{5}}s(\hat\mu)\ .
\end{align}
At $\mu=0$ and $T>0$ (and hence $\hat{\mu}=0$), we have $f(0)=s(0)=1$ (see Fig.~\ref{fig:hat_fmu_smu}), resulting in \eqref{eq:BFSS_energy}. 
As we can see from Fig.~\ref{fig:energies_comparison}, the finite-$\mu$ correction is rather small at $\mu\lesssim 1$.

\subsection{Estimation of further corrections} \label{sec:Corrections}

Taking the large-$N$ limit results in a classical description of supergravity at low temperatures. 
However, at intermediate and high temperatures, although we may not have finite-$N$ corrections, we certainly have the $\alpha'$-corrections. 

Studying intermediate and high temperatures, the authors of  Ref.~\cite{Berkowitz:2016jlq} obtained a few coefficients which are responsible for $\alpha'$-corrections in the BFSS matrix model. 
In particular, these include the coefficients $a_1$ and $a_2$ in the energy expansion \eqref{eq:BFSS_Energy_expansion} estimated to be
\begin{align} \label{eq:estimates_a1a2}
a_1=-10.0\pm 0.4\quad,\quad a_2=5.8\pm0.5 \text{.} 
\end{align}
Furthermore, it was possible to obtain an estimate for $b_1$ as
\footnote{These estimations included the assumption that $a_0$ and $b_0$ agree with the analytical gravity predictions. 
	Such an assumption is sensible to estimate unknown parameters. 
	Dropping it increases the error bars, but leads to a consistent estimation.}
\begin{align}
b_1=-3.5\pm2.0 \text{.}
\end{align}

An important question at this point is whether there are further significant unknown corrections to $E(T,\mu)$. 
For example, one expects cross-terms of the type $\mu \alpha' $ or $\mu / N^2$ to exist. 
To give a plausible estimate, we collected the size of energy predicted by classical supergravity at $T=0.3$ with $\mu=0$ plus all known corrections in table~\ref{tab:SizeOfCorrections}. 

\begin{table}[h]
	\centering
	\begin{tabular}{l | l | l}
		{} & {} & Estimated value \\
		{Contribution to $E(T,\mu) / N^2$} & {expression} & ~~  at $T=0.3$ \\
		\hline
		Classical supergravity at $\mu=0$ & $a_0 T^{\frac{14}{5}}$  & $+0.255$ \\
		first order $\alpha'$-correction at $\mu=0$  &  $a_1T^{\frac{23}{5}}$ & $-0.039$\\
		second order $\alpha'$-correction at $\mu=0$  & $a_2 T^{\frac{29}{5}}$ & $+0.005$\\
		finite $\mu$ correction to classical supergravity   &  \eqref{eq:BMN_sugra_energy}  $-$ \eqref{eq:BFSS_energy}  & +0.008 ~ (at $\mu = 0.5$)\\ \hline \hline 
		Sum of all known contributions & \eqref{eq:BMN_Energy_expansion_large_N}  & $+0.228$ \\
	\end{tabular}
	\caption{Contributions to $E(T, \mu) / N^2$ in the large $N$ limit rounded to the third digit after the decimal point. 
		Uncertainties from the estimations of $a_1$ and $a_2$ were ignored due to their smallness and thus irrelevance to the argument.
		We used the values of $a_1$ and $a_2$ obtained in Ref.~\cite{Berkowitz:2016jlq} by fitting the simulation results at $T\ge 0.4$.}
	\label{tab:SizeOfCorrections}
\end{table}

It transpires that the $\alpha'$-corrections are quickly vanishing and that the next correction (third order) is expected to be of order $0.001$. 
Since the first $\alpha'$-correction is about $16\%$ of the classical supergravity contribution and the finite $\mu$ correction is about $3\%$, we expect the corrections involving $\mu \alpha' $ to be also not larger than of order $0.001$. 
The same argument holds for $\mu / N^2$ or higher corrections when considering finite $N$. 
Since we have error bars to the order of $0.01$ in the simulations, these further corrections are insignificant for comparison. 

In conclusion, we argued that we should expect excellent agreement with simulations if we use the classical supergravity analysis for finite $\mu$~\cite{Costa:2014wya}, while using the $\alpha'$-corrections of first and second order, i.e. $a_1$ and $a_2$, as well as $b_1$ estimated in Ref.~\cite{Berkowitz:2016jlq}, which correspond to $\mu=0$. 
In other words, 
\begin{align}\label{eq:BMN_Energy_expansion}
\frac{E(T,\mu)}{N^2} \approx \frac{E(T,\mu)\Big|_{\rm sugra}+a_1 T^{\frac{23}{5}}+a_2 T^{\frac{29}{5}}}{N^0}+\frac{b_0 T^{\frac{2}{5}}+b_1 T^{\frac{11}{5}}+\cdots}{N^2}+\mathcal{O}(N^{-4}),
\end{align}   
where $E(T,\mu)\Big|_{\rm sugra}$ is given by \eqref{eq:BMN_sugra_energy}. 
In the large-$N$ limit and at very low temperatures we can assume the energy to be given by equation
\begin{align} \label{eq:BMN_Energy_expansion_large_N}
\frac{E(T,\mu)}{N^2} \approx \frac{E(T,\mu)\Big|_{\rm sugra}+a_1 T^{\frac{23}{5}}+a_2 T^{\frac{29}{5}}}{N^0}.
\end{align}

\section{Lattice setup}\label{sec:lattice-regularization}
The action is the same as the one used in Ref.~\cite{Berkowitz:2016jlq}, except that also the deformation terms are added (see also Ref.~\cite{Bergner:2021goh}).

\subsection{Gauge fixing}
The action of the BMN matrix model given in \eqref{eq:BMN_action} is invariant under the $SU(N)$ gauge transformation.
We take the static diagonal gauge,
\begin{eqnarray}
A_t=\frac{1}{\beta}\cdot{\rm diag}(\alpha_1,\cdots,\alpha_N),
\qquad
-\pi<\alpha_i\le\pi\ .
\end{eqnarray}
Associated with this gauge fixing, we add the Faddeev-Popov term defined by 
\begin{eqnarray}
S_{\rm F.P.}
&= &
-
\sum_{i<j}2\log\left|\sin\left(\frac{\alpha_i-\alpha_j}{2}\right)\right|
\label{eq:Faddeev-Popov}
\end{eqnarray}
to the action.

\subsection{Lattice action}

We regularized the gauge-fixed continuum theory by introducing a lattice with $L$ sites and spacing $a$. 
The time parameter $t$ takes the discrete values $t=a,2a,\cdots,La=\beta$. 
Breaking the action \eqref{eq:BMN_action} into the bosonic part $S_{\rm b}$, the fermionic part $S_{\rm f}$, the Faddeev-Popov term $S_{\rm F.P.}$ and the mass deformation parts $\Delta S_{\rm b}$ and $\Delta S_{\rm f}$, the respective lattice action is
\begin{eqnarray}
S_{\rm b}
&= &
\frac{N}{2a}\sum_{t}\sum_{I=1}^9{\rm Tr}\left(D_+X_I(t)\right)^2
-
\frac{Na}{4}\sum_t\sum_{I,J=1}^9{\rm Tr}[X_I(t),X_J(t)]^2.
\end{eqnarray}
\begin{eqnarray}
S_{\rm f}
=
iN\sum_{t}\Tr\bar{\psi}(t)
\left(
\begin{array}{cc}
0 & D_+\\
D_- & 0
\end{array}
\right)
\psi(t)
-
aN\sum_{t}\sum_{I=1}^9\bar{\psi}(t)\Gamma^I[X_I(t),\psi(t)],
\end{eqnarray}
\begin{align}
\Delta S_{\rm b}
=
aN\sum_{t}\Tr\left\{
\frac{\mu^2}{2}\sum_{i=1}^3X_i(t)^2
+
\frac{\mu^2}{8}\sum_{a=4}^9X_a(t)^2
+
i\sum_{i,j,k=1}^3\mu\epsilon^{ijk}X_i(t)X_j(t)X_k(t)
\right\},
\end{align}
and
\begin{eqnarray}
\Delta S_{\rm f}
=
\frac{3i\mu}{4}\cdot aN\sum_{t}\Tr\left(
\bar{\psi}(t)\gamma^{123}\psi(t)
\right),
\end{eqnarray}
where
\begin{eqnarray}
D_\pm\psi(t)
\equiv \mp\frac{1}{2}U^2\psi(t\pm 2a)\left(U^\dagger\right)^2
\pm 2U\psi(t\pm a)U^\dagger
\mp\frac{3}{2}\psi(t)
=
aD_t\psi(t) + O(a^3).
\nonumber\\
\end{eqnarray}
Here, $U={\rm diag}(e^{i\alpha_1/L},e^{i\alpha_2/L}\cdots,e^{i\alpha_N/L})$,
$-\pi\le \alpha_i<\pi$.
The Faddeev-Popov term $S_{\rm F.P.}$ is given in \eqref{eq:Faddeev-Popov}.

This lattice action is studied by using the Hybrid Monte Carlo algorithm. 
A potential issue is the sign problem, i.e., the Pfaffian appearing after integrating out fermions can have a complex phase. 
In this work, we omit the phase and use the absolute value of the Pfaffian, following the preceding work~\cite{Anagnostopoulos:2007fw,Catterall:2008yz,Hanada:2008ez,Hanada:2008gy,Hanada:2013rga,Kadoh:2015mka,Filev:2015hia,Berkowitz:2016jlq}.

\subsection{Observables}\label{sec:observables}

The observables we will consider throughout the paper are the energy of the system $E$, the Polyakov loop $P$, the sum of traces of the matrices squared $R^2$, and the Myers term $M$. 
First, we define them in terms of the continuous theory and then we present their lattice counterparts. 

To write the energy in a simple form, we use the virial theorem $\langle K\rangle=\frac{1}{2}\langle \sum\phi\frac{\partial V}{\partial\phi}\rangle$, where $K$ and $V$ are the kinetic and potential energies, and $\phi$ are the dynamical fields. 
We can write the total energy as
\begin{align}
\nn\frac{E}{N^2}:=&\frac{1}{N\beta}\int_0^\beta dt \Tr\Big\{-\frac{3}{4}[X_M,X_N]^2+\mu^2\sum_{i=1}^3X_i^3+\frac{\mu^2}{4}\sum_{b=4}^9X_b^2+\frac{5i\mu}{2}\sum_{i,j,k=1}^3\epsilon^{ijk}X_iX_jX_k\\
&-\frac{3}{2}\bar\psi \gamma^M[X_M,\psi]+\frac{3i\mu}{4}\bar\psi\gamma^{123}\psi\Big\}\ .
\end{align}
The Polyakov loop is defined via
\begin{align}\label{eq:Polyakov}
P:= \frac{1}{N}\Tr\left(\mathcal{P} \exp\left(i\int_0^\beta A_t dt\right)\right),
\end{align}
where $\mathcal{P}$ stands for path ordering. 
Another observable, which shows the stability of the simulation and potential runaway of a matrix eigenvalue, is defined as 
\begin{align}\label{eq:R2}
R^2:= \frac{1}{N\beta}\int_0^\beta dt\left(\sum_{I=1}^9 \Tr{(X_I)^2}\right).
\end{align}
The Myers term is given by 
\begin{align} 
M:= \frac{i}{3N\beta}\int_0^\beta dt\sum_{i,j,k}^3\epsilon_{ijk}\Tr X^iX^jX^k,
\end{align} 
and controls essentially the size of the fuzzy sphere background for the BMN model, while it is absent for the BFSS model.  

To obtain the lattice counterparts of these quantities, we just have to replace the integrals with the sums over the lattice points $t=a,2a,\cdots, La$. 
The energy is 
\begin{align}
\nn\frac{E}{N^2}=&\frac{a}{N\beta}\sum_t \Tr\Big\{-\frac{3}{4}[X_M(t),X_N(t)]^2+\mu^2\sum_{i=1}^3X_i^3(t)+\frac{\mu^2}{4}\sum_{b=4}^9X_b^2(t)\\
+&\frac{5i\mu}{2}\sum_{i,j,k=1}^3\epsilon^{ijk}X_i(t)X_j(t)X_k(t)
-\frac{3}{2}\bar\psi(t) \gamma^M[X_M(t),\psi(t)]+\frac{3i\mu}{4}\bar\psi(t)\gamma^{123}\psi(t)\Big\}\ .
\end{align}  
Because we use the static diagonal gauge, the Polyakov loop is defined by
\begin{align}
P=
\frac{1}{N}\sum_{j=1}^N e^{i\alpha_j}\ . 
\end{align}
The lattice counterparts of $R^2$ and $M$ are given by 
\begin{align}
R^2=\frac{a}{N\beta}\sum_{t}\sum_{I=1}^9
\Tr\left[X_I(t)\right]^2
\end{align}
and
\begin{align}
M=\frac{ia}{3N\beta}\sum_t\sum_{i,j,k=1}^3\epsilon_{ijk}\Tr X^i(t)X^j(t)X^k(t)\ .
\end{align}

\section{Simulation results}\label{sec:numerical_analysis} 
In this section, we present the numerical analysis and our main results. We will set $\lambda=1$. 
In Sec.~\ref{sec:NLmuT}, we explain the values of $N, L, \mu$ and $T$ we use for the simulations. Our target is the BFSS limit ($\mu=0$) at sufficiently low temperature, say $T=0.3$, and to take large-$N$ and continuum limit. We choose optimal values of $N, L$ and $\mu$ that enable us to achieve this goal with a smaller computational cost. For $T=0.3$, we will use $\mu=0.5$, $N\ge 10$ and $L\ge 24$, based on the reasons explained in Sec.~\ref{sec:NLmuT}. 
In Sec.~\ref{sec:PrecisionMeasurement}, we perform a detailed study of the large $N$ and continuum limit for $T=0.3$, $\mu=0.5$. 
In Sec.~\ref{sec:Tbigger}, we extend our study to various higher temperatures up to $T=0.8$. Section \ref{sec:Tsmaller} contains results about lower temperatures. We compare to a limited set of very large $N$ simulations at $\mu=0$ in section \ref{sec:ComparisonBFSS}. Finally, we collect our results in Sec.~\ref{sec:EvsT}, show the energy vs temperature plot, compare it to previous investigations, and provide improved estimates for the coefficients $a_1$ and $a_2$. 

To get precise results from simulations is quite challenging.  To give a rough estimate, let us have a look in Fig.~\ref{fig:DivergencesN} where we show the Monte Carlo histories and focus on the bottom right picture with parameters $N=16$, $L=30$, $T=0.3$ and $\mu=0.5$. To generate this particular picture we simulated in a cluster using 384 cores for roughly 18 days to produce these particular configuration points (roughly 8000 Monte Carlo trajectories). This simulation leads to one particular point out of 46 used in Fig.~\ref{fig:LargeNST03} to produce the precision result for $T=0.3$. Furthermore, the latter temperature is only one out of 6 points shown in Fig.~\ref{fig:EvsT_all_points}.  

\subsection{Appropriate choices of \texorpdfstring{$N$}{N}, \texorpdfstring{$L$}{L}, \texorpdfstring{$\mu$}{μ}, and \texorpdfstring{$T$}{T}}\label{sec:NLmuT}
In this paper, we are interested in the limit of large $N$ ($N=\infty$), continuum ($L=\infty$), BFSS ($\mu=0$) and strong coupling ($T\to 0$).  
Below, we explain the range of those parameters where the corrections are small and under control. 
\subsubsection{\texorpdfstring{$N$}{N} and flat direction}
As we saw in Sec.~\ref{sec:flat-direction}, the BFSS matrix model suffers from the flat direction problem, which becomes worse as temperature decreases in the deconfined phase. Due to this obstruction, it was not possible to simulate below $T=0.4$ in Ref.~\cite{Berkowitz:2016jlq}, or below $T=0.375$ in Ref.~\cite{Kadoh:2015mka}.\footnote{At the regularized level, the severeness of the instability can depend on the details of the regularization scheme. 
} Simulations going well below this temperature were either in the confined phase, which ameliorates the problem \cite{Bergner:2021goh}, or using constraints to prevent divergence of the matrix size $R^2$~\cite{Hanada:2013rga}. 
In particular, a minimal $N$ of 24 (resp., 32) was necessary in Ref.~\cite{Berkowitz:2016jlq} at $T=0.4$ (resp., in Ref.~\cite{Kadoh:2015mka} at $T=0.375$) to achieve stable simulations. 

\begin{figure}[ht!]
	\centering 
	\includegraphics[scale=0.43]{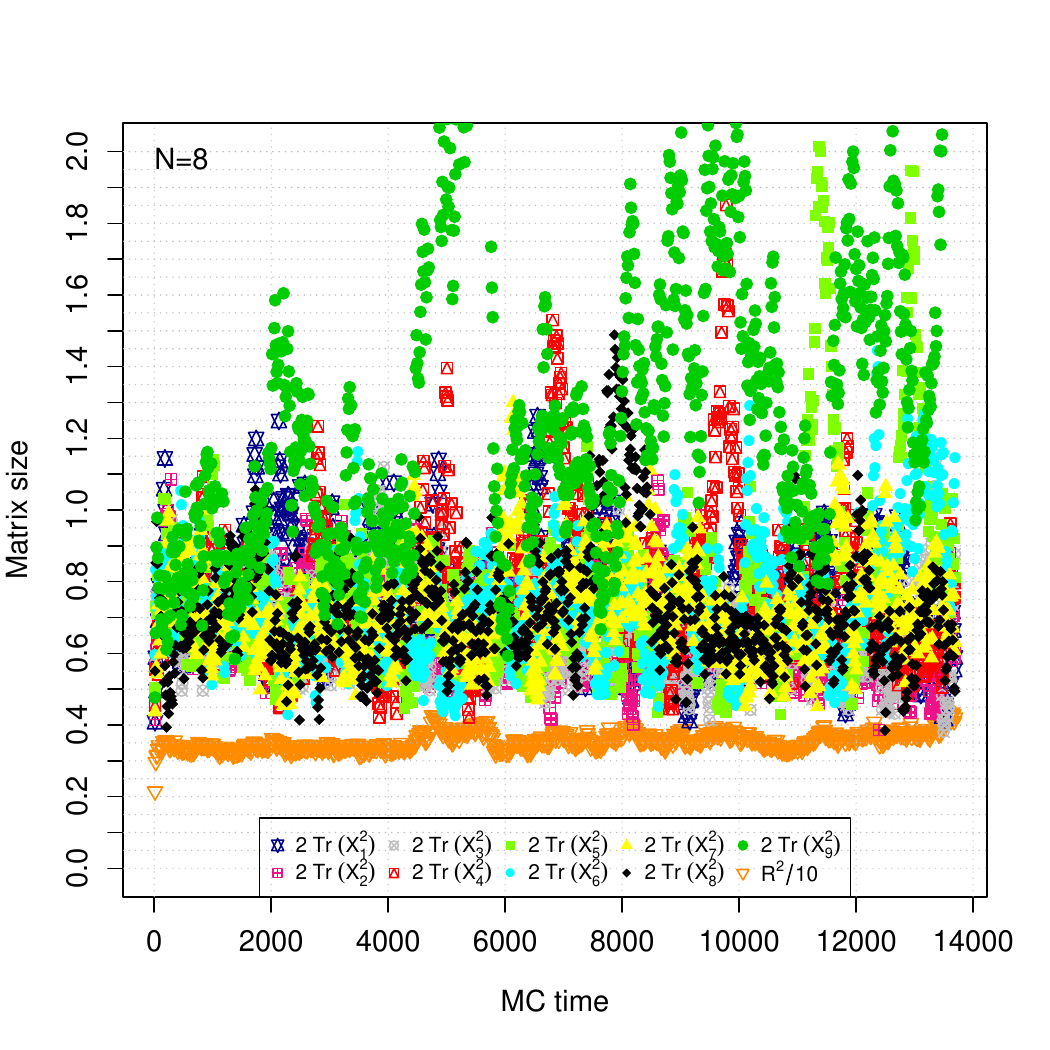}
	\includegraphics[scale=0.43]{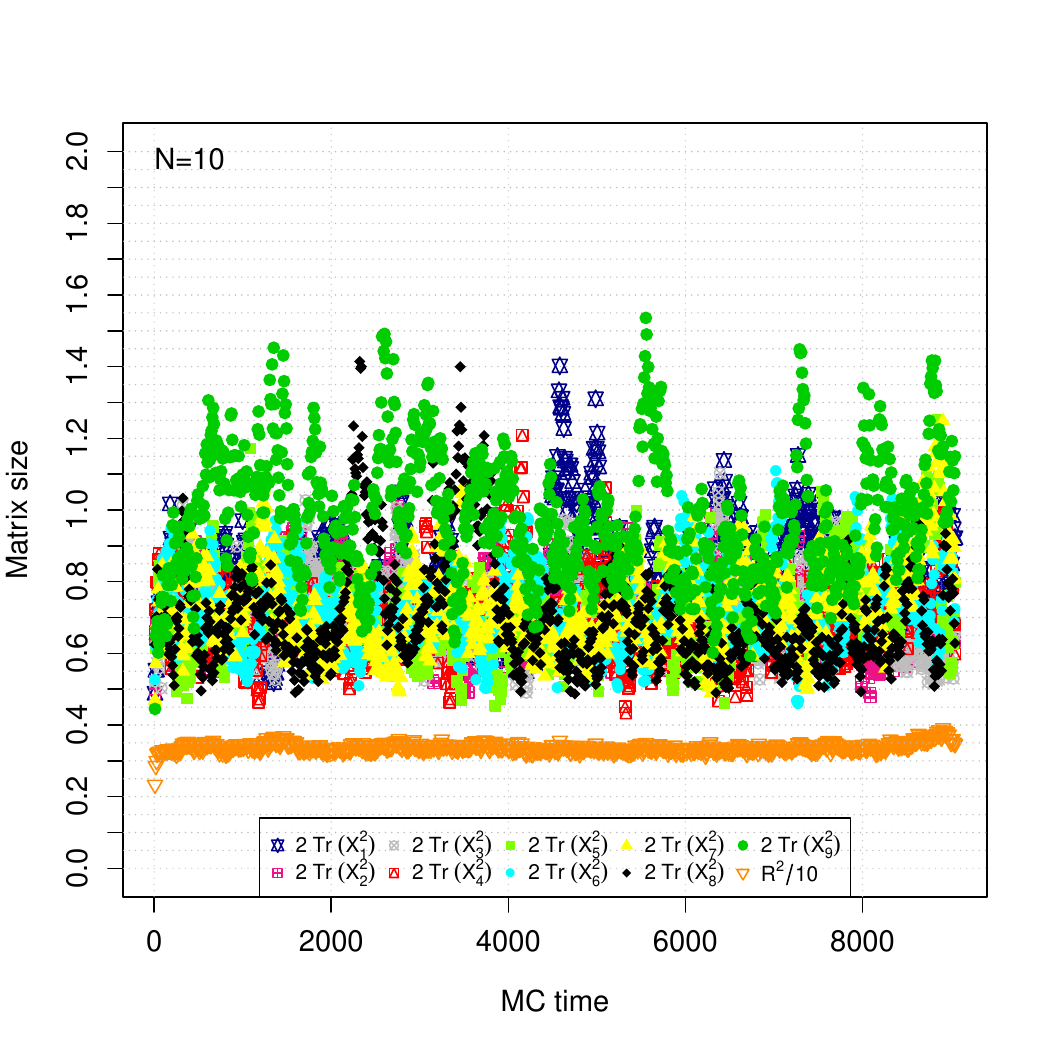} \\
	\includegraphics[scale=0.43]{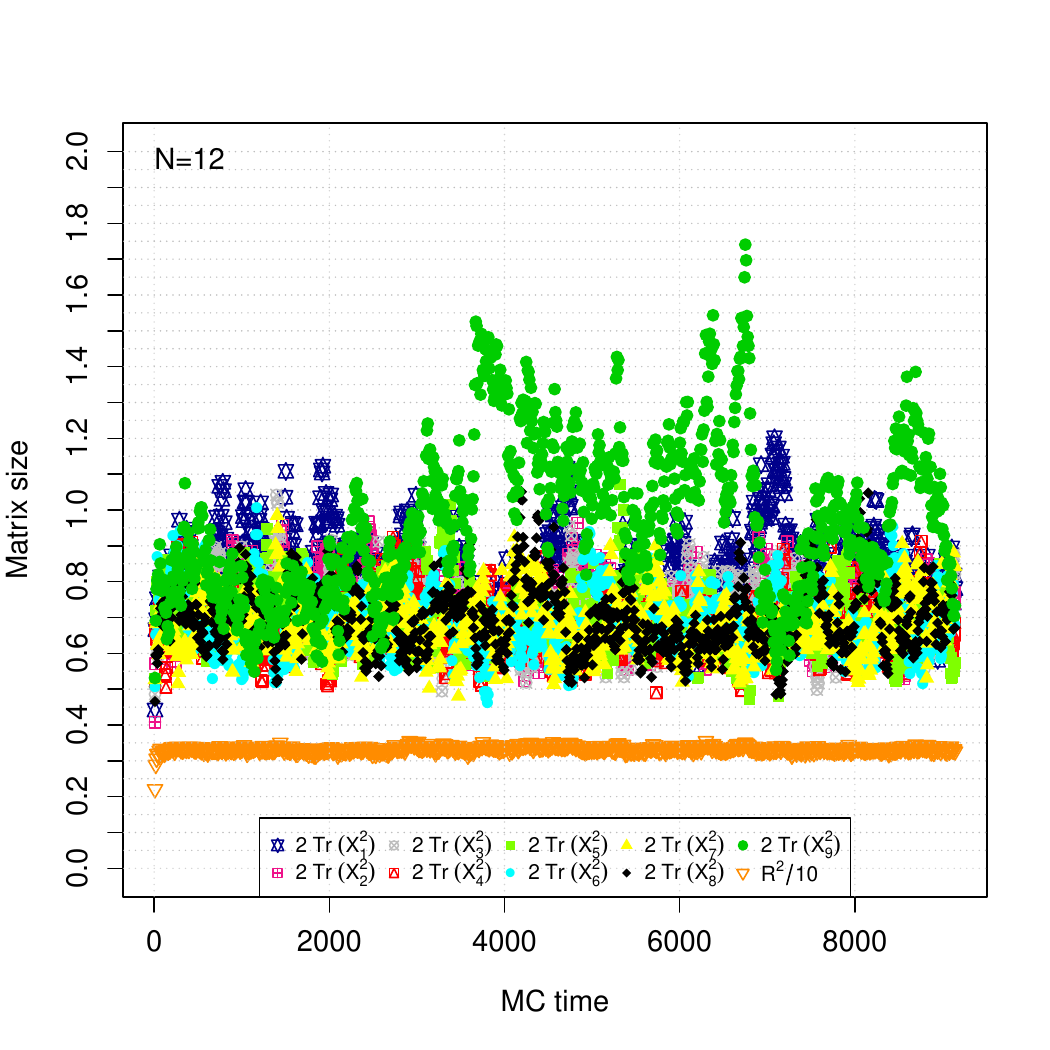}
	\includegraphics[scale=0.43]{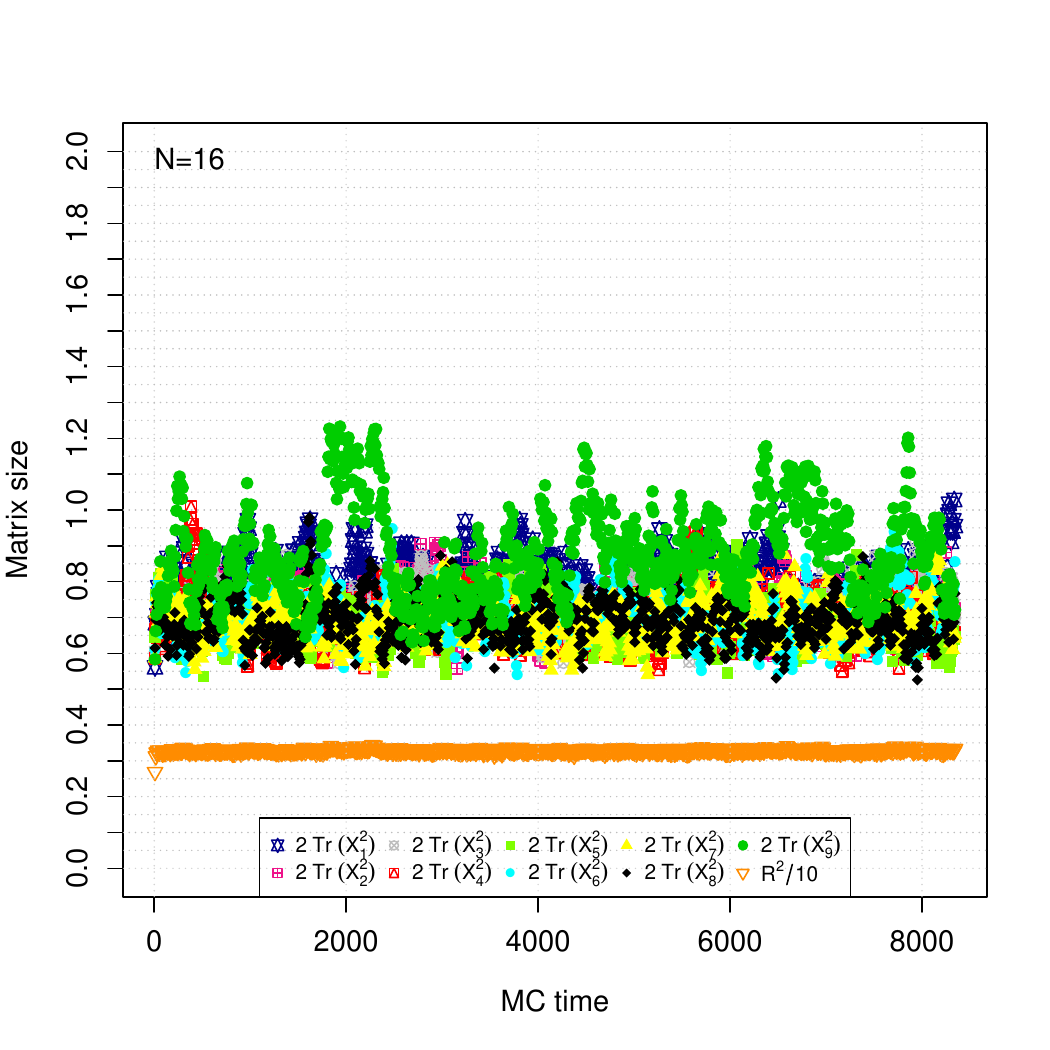} 
	\caption{Monte Carlo histories with $T=0.3$, $\mu=0.5$, $L=30$, $N=8,10,12,16$. We observe temporary increases in ${\rm Tr}X^2$ more frequently and strongly at lower $N$. Some green data points for $N=8$ are above the plotting range.}
	\label{fig:DivergencesN}
\end{figure}

In our simulations of the BMN model at $\mu=0.5$, starting at $N=8$, we found that the instabilities mostly disappear, although occasional excursions to the large-$R^2$ region are seen at $N = 8, 10, 12$. This ceases to be the case at about $N=16$, see Fig.~\ref{fig:DivergencesN}. We conclude that care needs to be taken when including lower $N$ in the analysis, in particular when considering the matrix sizes, as it is a priori unclear to which extent divergences in the matrices affect the true physical results.

\subsubsection{\texorpdfstring{$L$}{L} and continuum limit}

\begin{figure}[ht!]
	\centering 
	\includegraphics[scale=0.65]{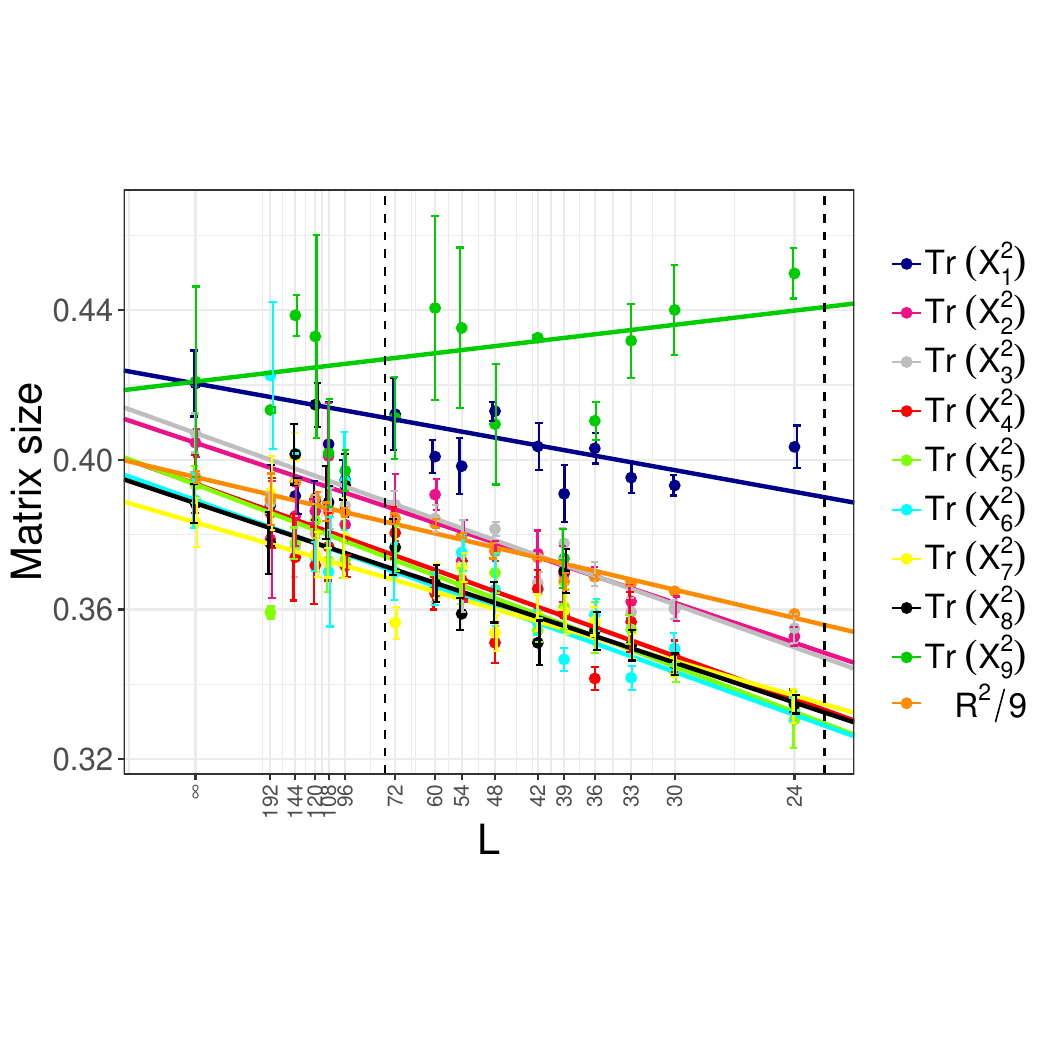}
	\caption{Sizes of the nine matrices and their average as a function of $L$, plotted with $1/L$ scaling, for $N=16$, $T=0.3$, $\mu=0.5$. We observe agreement in the continuum limit, showing that the observed symmetry breaking is a lattice artifact. The extrapolation is linear using data points within the dashed lines. $L > 72$ is omitted due to low statistics and a likely underestimation of the error bars.}
	\label{fig:SymmetryBreakingContinuum}
\end{figure}

As can be seen from Fig.~\ref{fig:DivergencesN}, the ninth matrix appears to have a larger expectation value than the other matrices, hinting at an apparent symmetry breaking of the SO$(9)$ symmetry acting on the matrices. This symmetry breaking is a lattice artifact that disappears in the continuum limit (see also Ref.~\cite{Schaich:2022duk} for the same conclusion based on a different lattice action). We verify this by taking the continuum limit of the individual matrix expectation values in Fig.~\ref{fig:SymmetryBreakingContinuum} at the example of $N=16$. Other values of $T$, $\mu$, and $N$ lead to the same conclusion. We observed that when going well below $24$ lattice points at $T \approx 0.3$, this effect is much stronger and an approximately linear interpolation as in Fig.~\ref{fig:SymmetryBreakingContinuum} is not possible anymore. Hence, we restricted our simulations to $L\ge 24$ to avoid possible non-trivial issues associated with the continuum extrapolation. 

\subsubsection{\texorpdfstring{$\mu,T$}{μ,T} and BFSS limit}

There are two reasons to take $\mu$ as small as possible. First, for the comparison with the gravity predictions~\cite{Costa:2014wya} to make sense, we need to take both $\mu$ and $T$ to be small, and furthermore, the combination $\hat\mu\equiv \frac{7\mu}{4\pi T}$ to be small. Even when $T$ is not small, if $\mu$ is sufficiently small the $\alpha'$-corrections studied for $\mu=0$ in the past can be reproduced.\footnote{
	We expect that the correction is small because the leading correction is of order $\mu^2$. This is the feature of the matrix model and hence valid including the $\alpha'$-corrections on the gravity side. 
} 
Next, decreasing $\mu$ lowers the temperature at which the transition to the confined phase takes place~\cite{Costa:2014wya,Bergner:2021goh} so that lower temperatures can be studied. As we saw in section \ref{sec:Corrections}, the finite-$\mu$ correction to the energy is expected to be very small at the target temperatures $T \simeq 0.3$ already at $\mu \simeq 0.5$. Additionally, it was shown in Ref.~\cite{Bergner:2021goh} that the deconfined phase exists for $\mu=0.5$ at $T \gtrsim 0.25$ at the values of $N$ considered in this paper.

\begin{figure}[ht!]
	\centering 
	\includegraphics[scale=0.43]{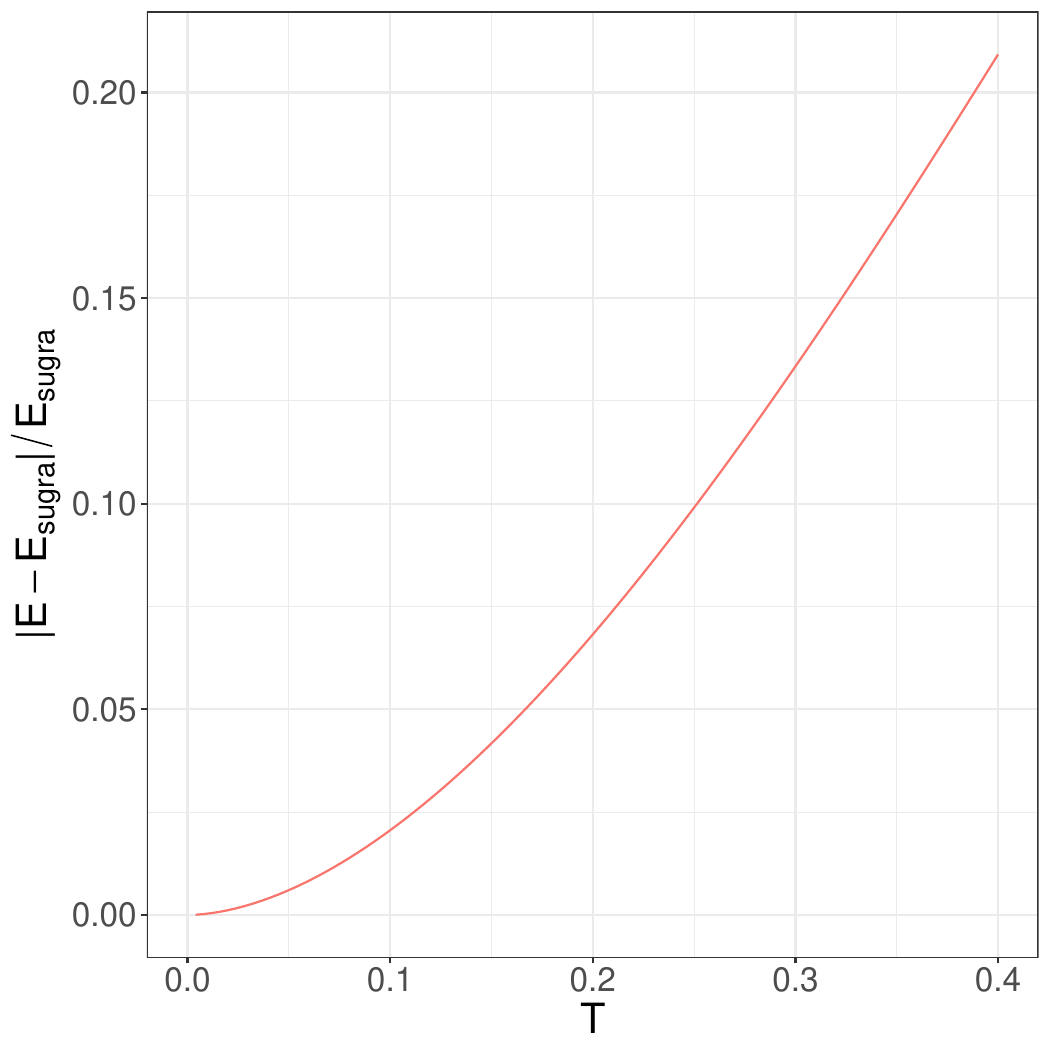}
	\includegraphics[scale=0.43]{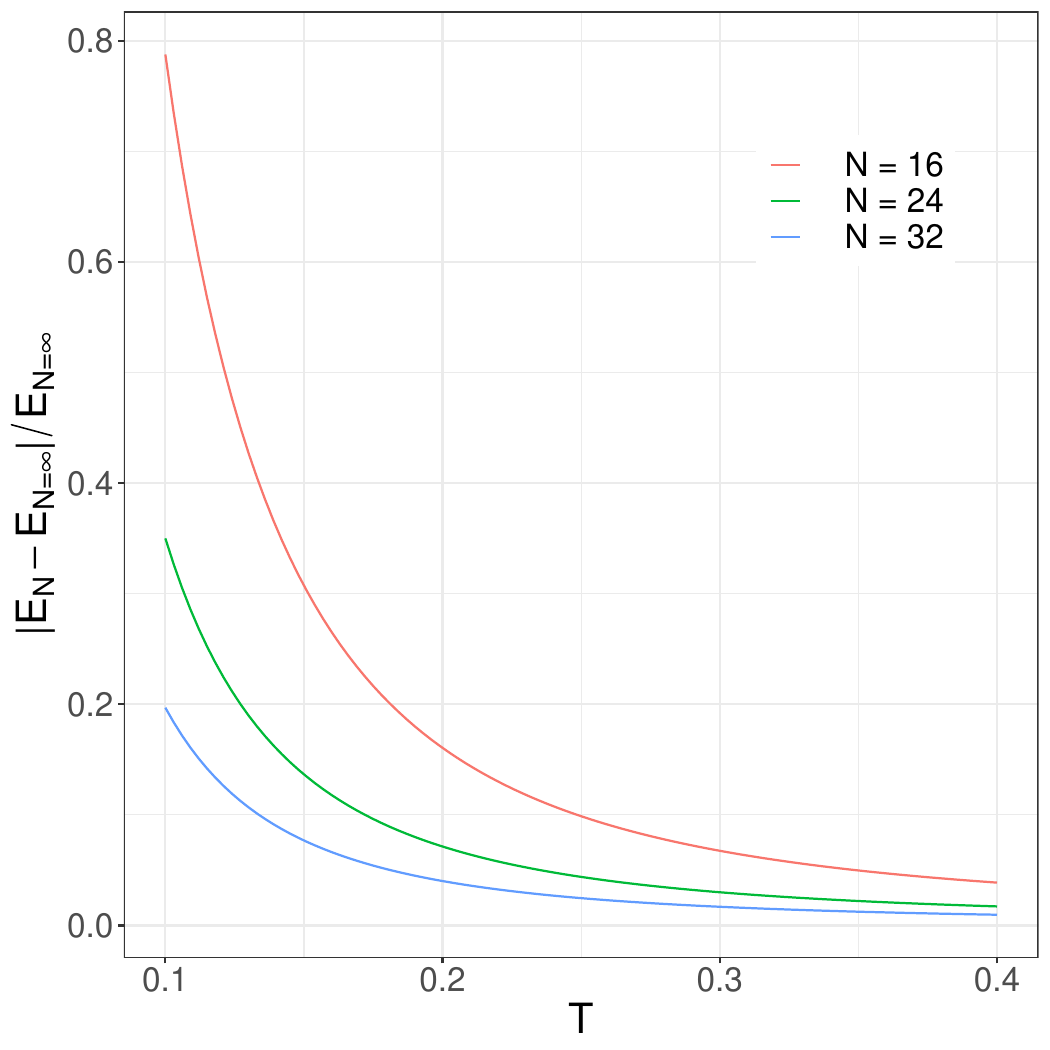}
	\caption{[Left] Relative size of the $\alpha'$-corrections in the BFSS model as a function of $T$. [Right] Relative size of the $1/N^2$ term in equation \eqref{eq:BMN_Energy_expansion} to the BFSS energy as a function of $T$ for various $N$.}
	\label{fig:AlphaprimeandfiniteNsize}
\end{figure}

Going below $T=0.25$ likely requires a much larger $N$, and we found in a preliminary analysis that $N=24$ is probably not enough to reach $T=0.2$ at $\mu=0.5$, as the Monte Carlo chain always quickly tunnelled to the confined phase\footnote{In the large $N$ limit, the gravity analysis of \cite{Costa:2014wya} predicts, see eq. \eqref{eq:TcLargeN}, $T_c = 0.159$ for $\mu=0.5$, so that we expect to reach down as far as this temperature at sufficiently large $N$.}. Going to even lower $T$ would be desirable as one gets closer to the classical gravity regime so that no simulation-informed estimate of the $\alpha'$-corrections is necessary to establish the agreement with supergravity. Fig.~\ref{fig:AlphaprimeandfiniteNsize} highlights the relative size of the $\alpha'$-corrections estimated by the fit of the matrix model simulation results in Ref.~\cite{Berkowitz:2016jlq}, showing that for $T=0.25$ we are only $10\%$ away from the supergravity limit. Fig.~\ref{fig:AlphaprimeandfiniteNsize} however also shows that the relative size of the finite $N$ corrections rises as $T$ gets smaller, indicating that the perturbation expansion becomes unreliable at too small $N$, consistent with the above discussion that the deconfined phase requires large $N$ to exist at low $T$. 

\begin{figure}[ht!]
	\centering 
	\includegraphics[scale=0.45]{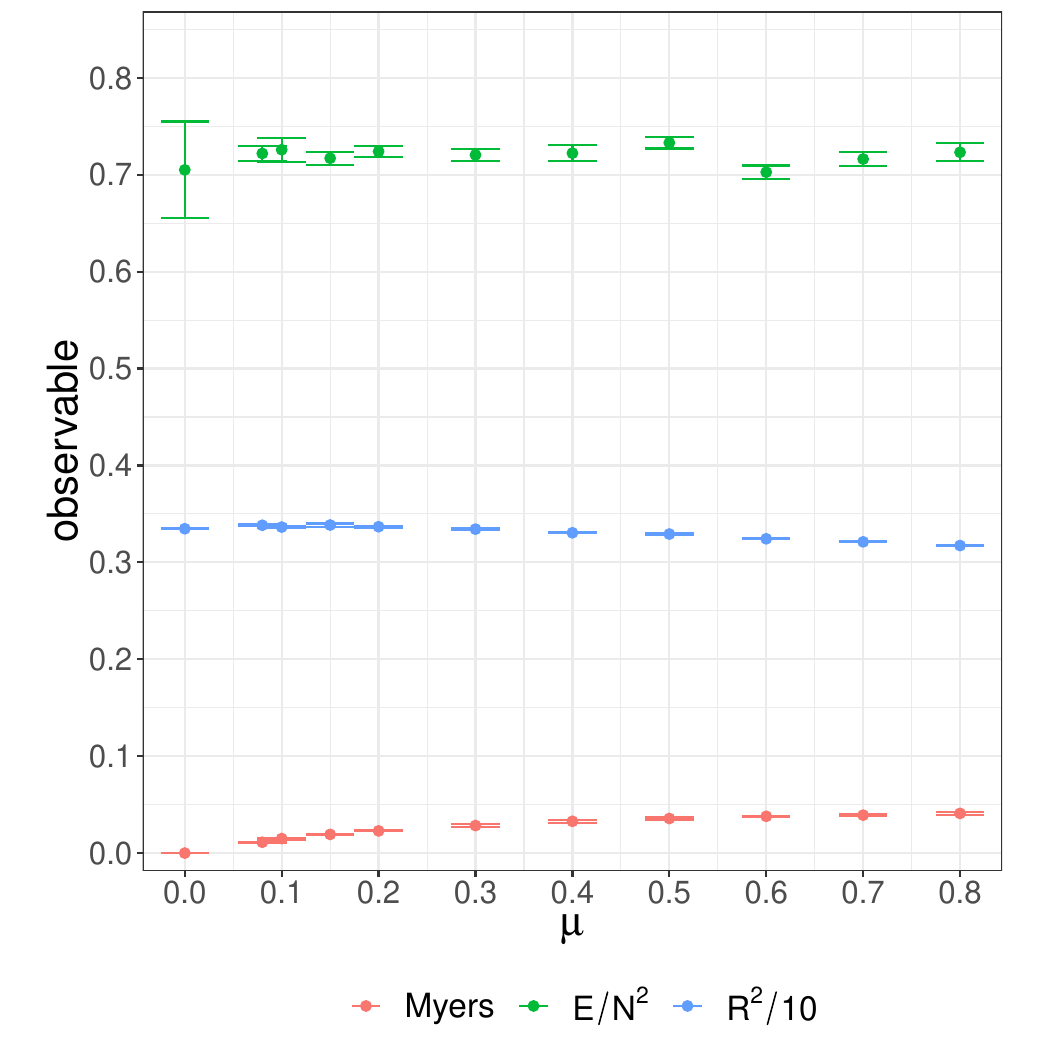}
	\caption{The dependence of a few observables on $\mu$ at $T=0.4, N=16, L=24$. Data points at $\mu=0$ for the same parameters are from \cite{Berkowitz:2016jlq}. We observe that the BMN energy for finite $\mu$ is contained within the error bars of the BFSS energy and also the other observables approach the BFSS values for $\mu \rightarrow 0$.  }
	\label{fig:mu_dependence}
\end{figure}

At moderately high temperatures where the flat direction is better under control, we can study the $\mu$-dependence. 
As discussed in Sec.~\ref{sec:Corrections}, we expect the finite-$\mu$ corrections to the energy to be small. In order to verify this, we simulated several values of $\mu\leq 0.8$ at $T=0.4$, $N=16$, $L=24$, where simulation results for $\mu = 0$ are available for direct comparison \cite{Berkowitz:2016jlq}. It is clear from Fig.~\ref{fig:mu_dependence} that the $\mu$-dependence is very small and within the error bars from the $\mu=0$ measurement for the whole range of $\mu\leq 0.8$.

\begin{figure}[ht!]
	\centering 
	\includegraphics[scale=0.45]{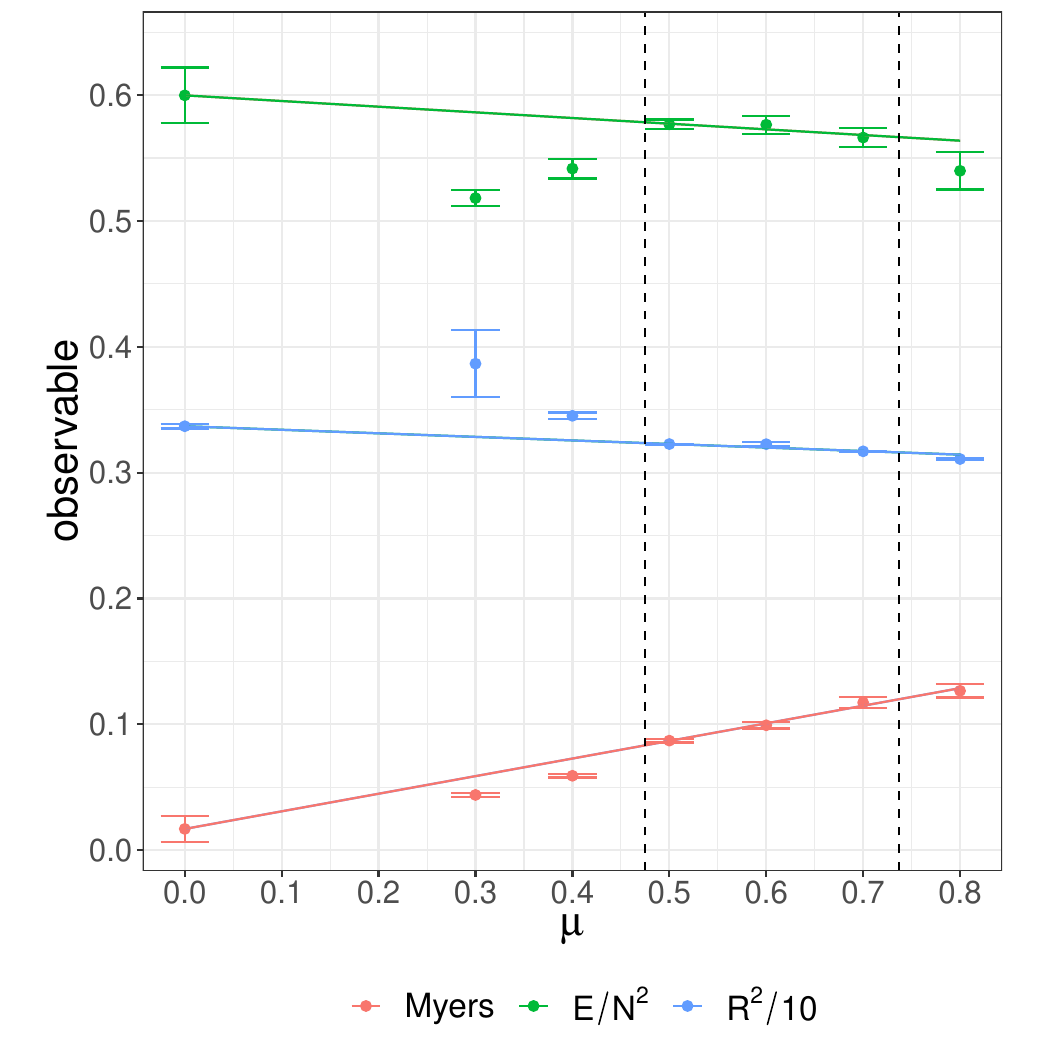}
	\caption{The expectation values of a few observables at $T=0.3$, $N=16$, $L=24$ are plotted against $\mu$. Above and below $\mu=0.5$, the behavior of $R^2$ is different. 
		A quick increase of $R^2$ at $\mu<0.5$ is associated with instability associated with the flat direction in the BFSS limit. Other observables change the behavior, too, once the instability sets in. The data points at $\mu=0$ are the linear extrapolations for values of $\mu$ in $0.5\leq\mu\leq0.7$. The linear extrapolations are only used to indicate, in accordance with figure 8, that we see a negligible slope for $E$ as well as to highlight the diverging behavior for $\mu < 0.5$ that sets in at lower temperatures. The $\mu\to 0$ limit was studied in a different setting in Ref.~\cite{Pateloudis:2022GvsU} and a quadratic extrapolation was used for observables such as $E/N^2$ and $R^2/N^2$, however, we are agnostic about odd or even preference of the $\mu$ extrapolations.}
	\label{fig:SmallMuDivergence}
\end{figure}

At low temperatures, it is problematic to take $\mu$ too small, because the flat direction is a more serious issue there. We simulated several $\mu$ for $T=0.3$ at $N=16$ and $L=24$, see Fig.~\ref{fig:SmallMuDivergence}.  We find that $\mu = 0.5$ sufficiently stabilizes the simulation, i.e. $R^2$ does not show any sign of divergence. The instability sets in at around $\mu=0.4$ and already at $\mu=0.3$ we see a significant increase in $R^2$ above the expected (non-divergent) BFSS value of about 3.3. 

Combining both estimates above, we find that simulating at $\mu=0.5$ is most feasible, as this value satisfies all requirements. There does not seem to be any need to go much above $\mu=0.5$, nor the possibility to go much below the currently accessible values of $N$. 

As for the temperature, we encounter frequent transitions to the confined phase around $T=0.25$, so simulations would need to be either supplemented by constraints or frequently restarted to obtain sufficient statistics. For this reason, we chose to focus on $T=0.3$ for clean precision measurement and collected only limited data below this temperature. Some higher temperatures where simulations are much cheaper were included for comparison with the references that studied the BFSS model.

\subsection{Precision measurement at \texorpdfstring{$T=0.3$}{T=0.3}} \label{sec:PrecisionMeasurement}
\begin{figure}[ht!]
	\centering 
	\includegraphics[scale=0.43]{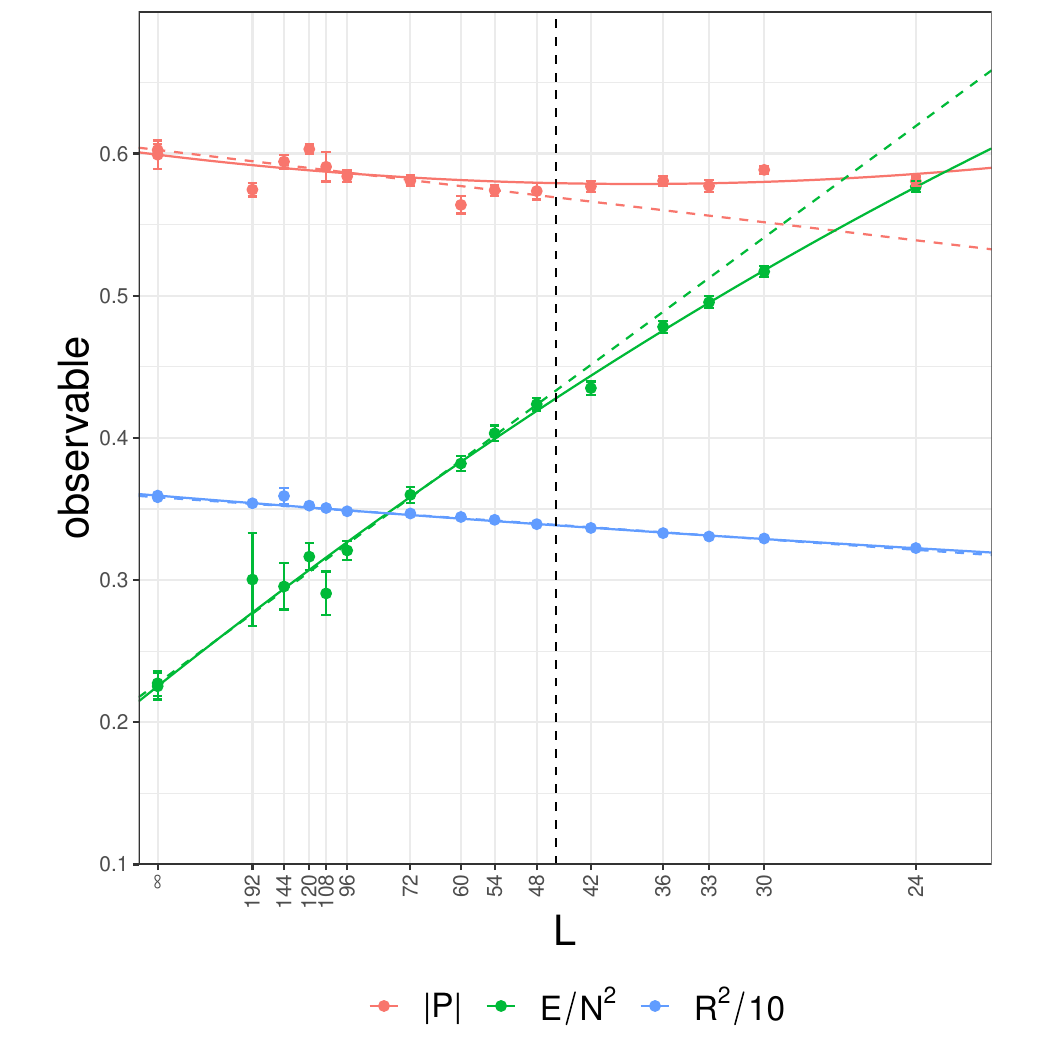}
	\includegraphics[scale=0.43]{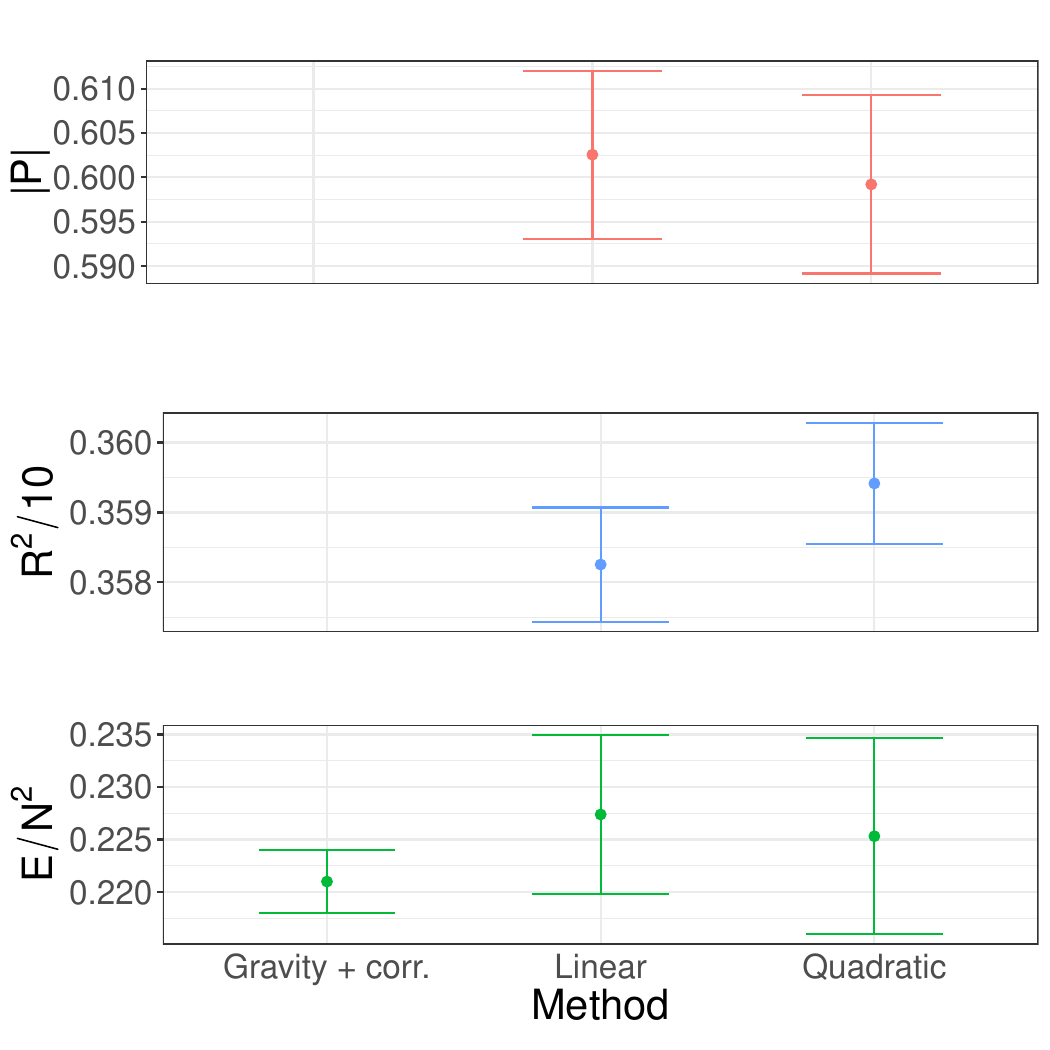}
	\caption{[Left] The extrapolation to the continuum limit ($L\to\infty$) for $T=0.3$, $N=16$, $\mu=0.5$. Solid lines are quadratic fits in $\frac{1}{L}$. Dashed lines are linear fits in $\frac{1}{L}$ that used the data points at the left of the dashed black line. These extrapolation ans{\"a}tze lead to consistent extrapolation values. 
		[Right] The continuum-extrapolated values are shown. For the energy, we also show the classical gravity prediction at $\mu=0.5$ including $\alpha'$ and finite $N$ corrections that were estimated using matrix model simulations. Specifically, we use $a_1, a_2, b_1$ as estimated in \cite{Berkowitz:2016jlq} from BFSS simulations at $T\geq 0.4$. Additionally, we use $\varepsilon_{2,0}$ as estimated in table \ref{tab:FitT0.3largeNSfreecNN}, which gives a significant contribution of $+0.007$. Error bars on the gravity result originate from uncertainties on the coefficients $a_1, a_2, b_1, \varepsilon_{2,0}$ and don't include an estimation of higher order corrections.}
	\label{fig:LargeSN16}
\end{figure}

In this subsection, we perform a detailed investigation of the large-$N$ and continuum limit for $T=0.3$. We also aim to understand the magnitude of finite-$N$ and finite-$L$ corrections in order to choose suitable fitting functions 
\begin{align}\label{eq:extrapolation_ansatz}
\frac{E(N,L)}{N^2}=\sum_{i,j=0}^k\frac{\varepsilon_{i,j}}{N^{2i}L^j}.
\end{align}
The largest influence on the energy originates from finite $L$ corrections. Due to the small temperature as compared to previous investigations such as Ref.~\cite{Berkowitz:2016jlq}, we expected to need quite a large $L$. Fig.~\ref{fig:LargeSN16} shows the extrapolation to the continuum limit ($L\to \infty$) for $T=0.3$, $N=16$, and $\mu=0.5$. It transpires that a quadratic fit in $1/L$ is necessary when including lattices below $L=48$ and also sufficient until $L=24$, while a linear fit in $1/L$ is sufficient for $L \ge 48$. When changing the temperature, we expect suitable fitting ranges to scale as $1/T$, i.e. lower temperatures require larger $L$. 

\begin{figure}[ht!]
	\centering 
	\includegraphics[scale=0.43]{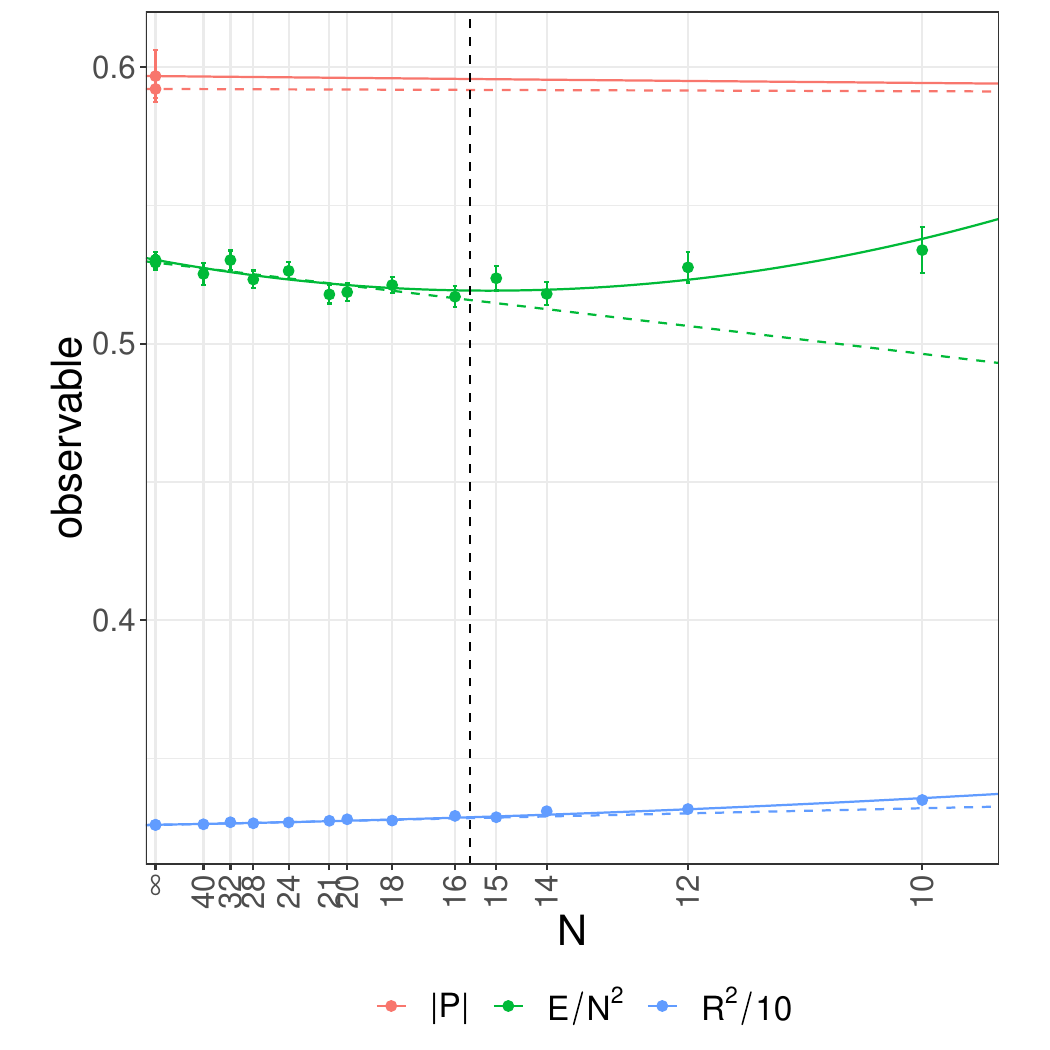}
	\includegraphics[scale=0.43]{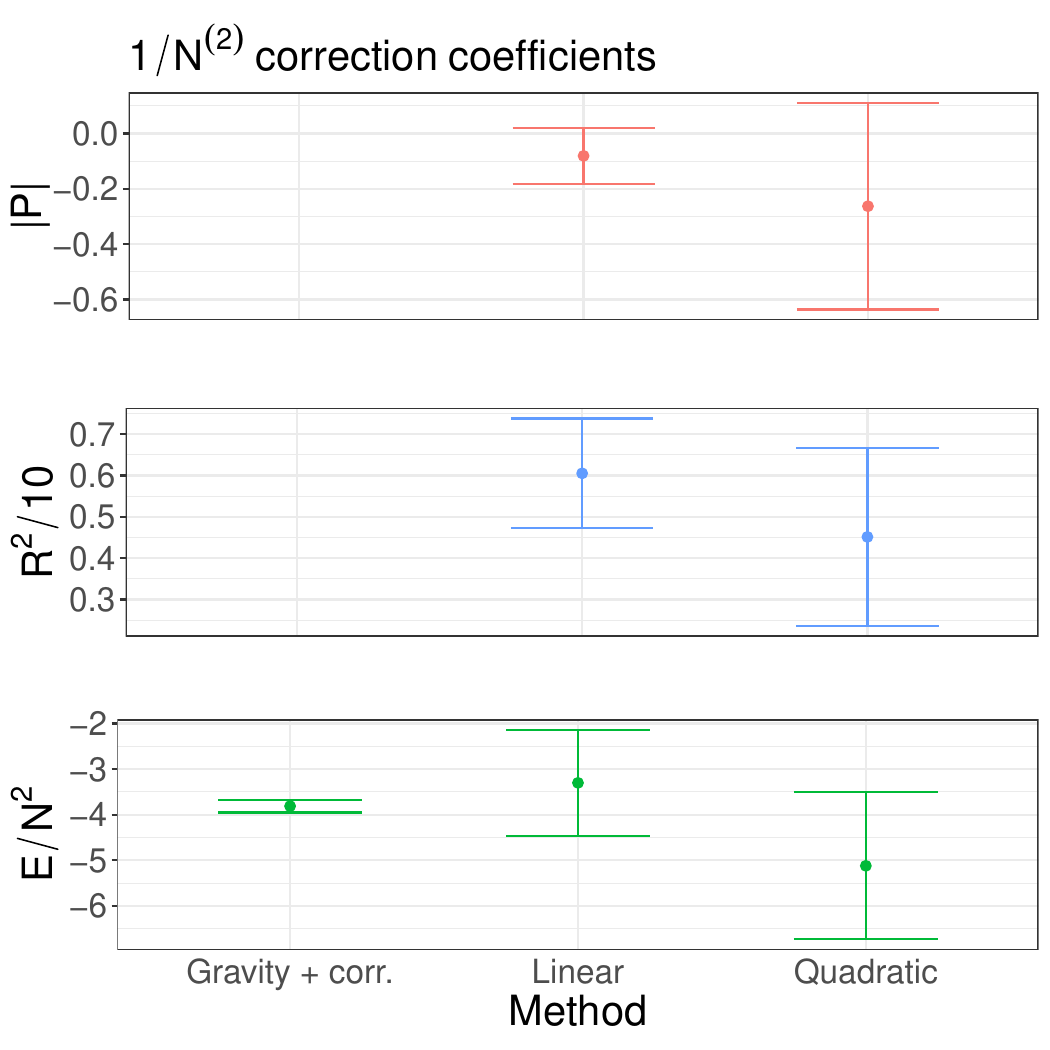}
	\caption{[Left] Large-$N$ extrapolation for $T=0.3$, $L=30$, $\mu=0.5$. The first axis scales as $1/N^2$. Solid lines are quadratic fits in $1/N^2$. Dashed lines are linear fits in $1/N^2$ that use the data points at the left of the dashed black line. Due to the $1/N$ scaling of corrections to $|P|$, the data points underlying the fit of $|P|$ are outside of the plotting range. 
		[Right] The coefficient $\varepsilon_{1,0}$ of the $1/N^2$ correction is obtained from the fit. These extrapolation ans{\"a}tze lead to consistent extrapolation values. For the energy, we also show the gravity prediction obtained in \cite{Berkowitz:2016jlq}, which includes a numerical estimate of the parameter $b_1$ from the fit of the BFSS model at $T\geq 0.4$. Error bars on the gravity result originate from uncertainties on the coefficient $b_1$ and don't include an estimation of higher order corrections.  }
	\label{fig:LargeNS30}
\end{figure}

Finite $N$ corrections are much smaller and also much harder to estimate precisely, as already observed for $\mu=0$~\cite{Berkowitz:2016jlq}. We chose to invest most effort in simulations at fixed $L=30$ due to the lower simulation cost as compared to larger $L$. Fig.~\ref{fig:LargeNS30} shows the behaviour of the observables as a function of $1/N^2$. We conclude that a quadratic fit in $1/N^2$ is necessary when including $N \geq 10$, while a linear fit seems sufficient for $N \geq 16$ to capture the trend at large $N$.

\begin{figure}[ht!]
	\centering 
	\includegraphics[trim={0cm 0 0 13mm}, clip, scale=0.7]{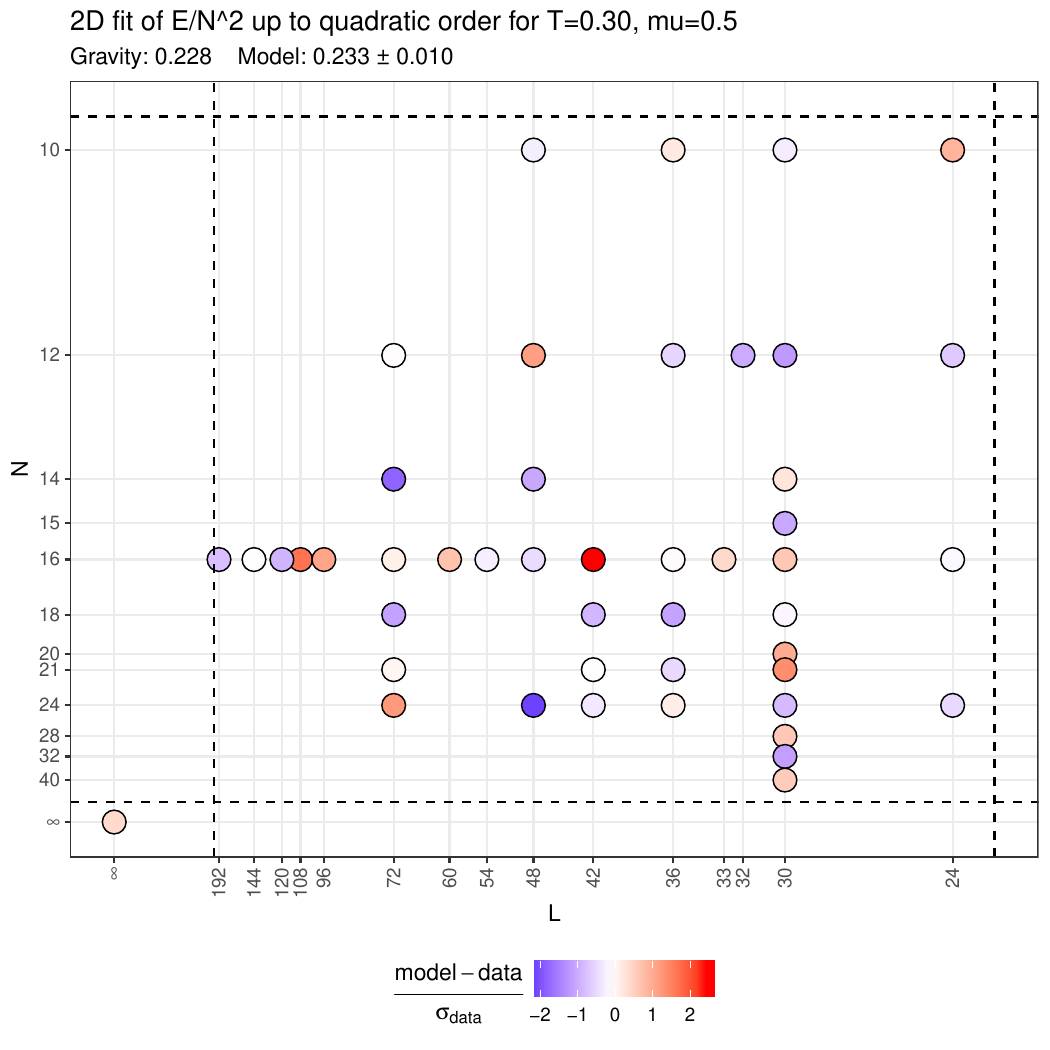}
	\caption{ Simultaneous large-$N$ and continuum extrapolation of the energy at $T=0.3$, $\mu=0.5$ using the ansatz \eqref{eq:quadratic_ansatz}. Datapoints within the dashed lines are included in the fit. Fitting results are summarized in table \ref{tab:FitT0.3largeNSfreecNN}. The colored circles in the figure encode the normalized deviation of the measurements from the fit. The absence of localized clusters of over- or underestimations provides a first indication of the suitability of the fitting ansatz. A quantitative statistical test is presented in figure \ref{fig:LargeNSKS}. The large $N$ continuum result is compared to the gravity prediction including estimates of $a_1, a_2$ from \cite{Berkowitz:2016jlq}.}
	\label{fig:LargeNST03}
\end{figure}

\begin{figure}[ht!]
	\centering 
	\includegraphics[scale=0.45]{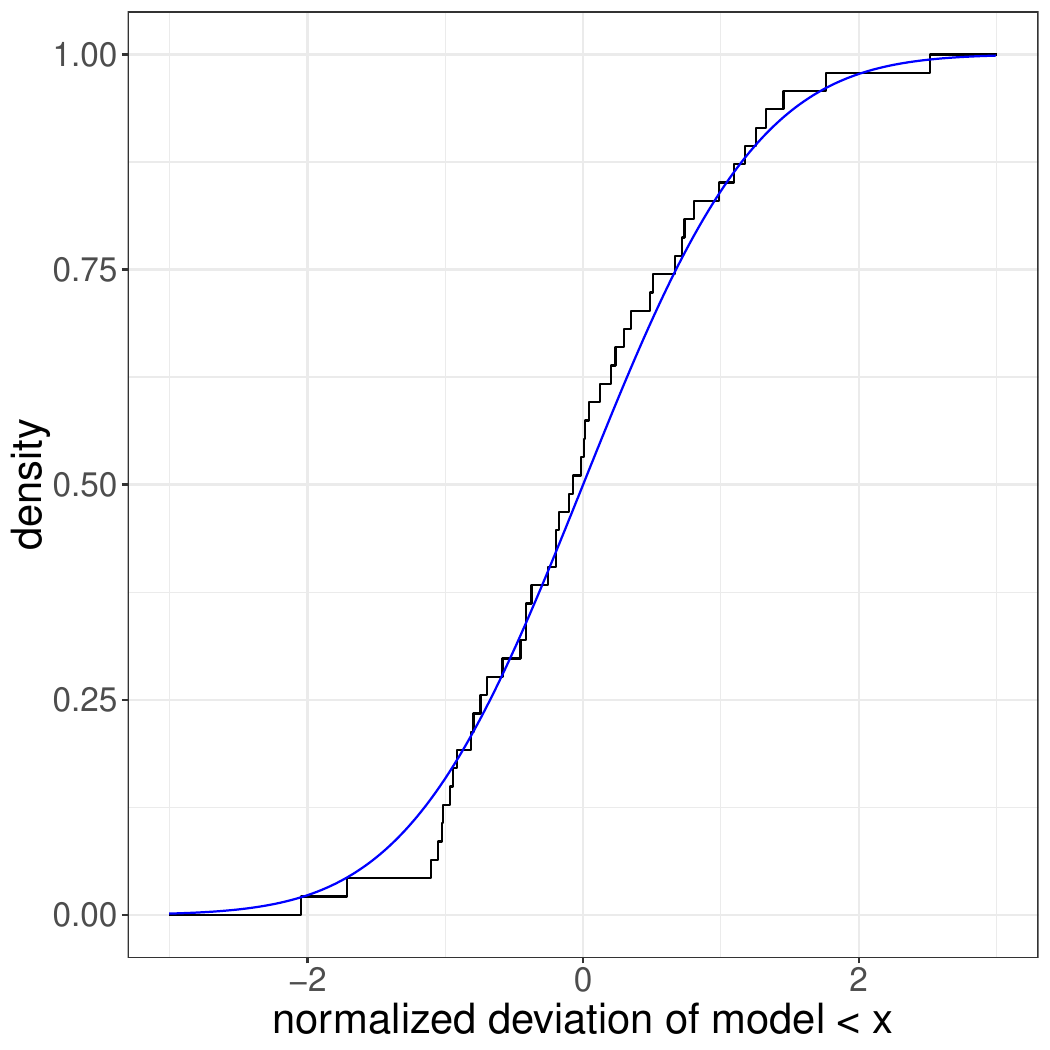}
	\caption{We plot the cumulative distribution function of the relative differences of model and measurements observed in figure \ref{fig:LargeNST03} in black along with the cumulative distribution function of the standard normal in blue. A maximal deviation between the cumulative distribution functions of $d_{\text{max}} =  0.092$ is observed. A Kolmogorov-Smirnov test (at $95\%$ with a critical value of 0.215 at 40 degrees of freedom) is successfully passed, giving faith in the suitability of the fitting ansatz as well as the estimation of error bars in the statistical analysis of the Monte Carlo data.  }
	\label{fig:LargeNSKS}
\end{figure}

Next, we performed a simultaneous large $N$ and continuum extrapolation using the most general ansatz up to quadratic order in $1/L$ and $1/N^2$ given by   
\begin{align}\label{eq:quadratic_ansatz}
\frac{E(N,L)}{N^2}=\varepsilon_{0,0}+\frac{\varepsilon_{1,0}}{N^2}+\frac{\varepsilon_{2,0}}{N^4}+\frac{\varepsilon_{1,1}}{L N^2}+\frac{\varepsilon_{0,1}}{L} +\frac{\varepsilon_{0,2}}{L^2} \text{.}
\end{align}
The result is presented in Fig.~\ref{fig:LargeNST03} and table \ref{tab:FitT0.3largeNSfreecNN}, showing excellent agreement with the gravity prediction plus expected corrections (finite $\mu$ and $\alpha'$) \footnote{Note that Sec.~\ref{sec:Corrections} uses $\alpha'$-corrections emanating not from gravitational analysis, since there is none, but from a numerical fit using matrix models \cite{Berkowitz:2016jlq}. It is merely an estimate for $\alpha'$-corrections. } explained in Sec.~\ref{sec:Corrections} within error bars. 
As a cross-check of our analysis, we perform a Kolmogorov-Smirnov test based on the fit with ansatz \eqref{eq:quadratic_ansatz} in figure \ref{fig:LargeNSKS}. The test shows very good agreement between the two cumulative distribution functions, indicating that a) the ansatz \eqref{eq:quadratic_ansatz} contains sufficiently many terms to accurately describe the measured data and that b) the estimation of the statistical error bars in the Monte Carlo simulations was accurate. Otherwise, we would have seen that a) the observed cumulative distribution function does not resemble a standard normal, or b) it resembles a stretched standard normal, i.e., with rescaled argument.

\begin{table}[ht]
	\centering
	\begin{tabular}{| l || l | l | l || l |}
		\hline
		coefficient & \multicolumn{3}{c||}{fit} & Estimate based on \\ 
		& value & error & t-value  & ~~~~ Ref.~ \cite{Berkowitz:2016jlq}  \\
		\hline
		\hline
		$\varepsilon_{0,0}$     & 0.232  &   0.01  &  24.5 &  0.228  \\ \hline
		$\varepsilon_{1,0}$   & -3.97   &  1.8  &  -2.21  &  -3.81 \\ \hline
		$\varepsilon_{2,0}$   & 481   &  124  & 3.88  &  unkown \\ \hline
		$\varepsilon_{0,1}$    & 10.6  &   0.54   & 19.7  & none \\ \hline
		$\varepsilon_{0,2}$ &-49.6  &  8.6 &  -5.75  & none \\ \hline
		$\varepsilon_{1,1}$ &-20.5  &  46.1 &  -0.45  & none \\ \hline
	\end{tabular}
	\caption{Simultaneous large-$N$ and continuum fit for $T=0.3$, $\mu=0.5$ with ansatz \eqref{eq:quadratic_ansatz}. Residual standard error: 0.986 on 40 degrees of freedom. Data includes unconstrained simulations at $N \geq 10$. }
	\label{tab:FitT0.3largeNSfreecNN}
\end{table}

\subsection{Higher temperatures \texorpdfstring{($T > 0.3$)}{(T>0.3)}} \label{sec:Tbigger}
For temperatures higher than $T=0.3$, we collected limited statistics at $\mu=0.5$ in order to compare with the BFSS results at $T\ge 0.4$~\cite{Berkowitz:2016jlq}. For the points $T=0.35$ and $T=0.4$ we are still using quadratic fits with lattice size $L=30,\cdots,144$. On the contrary, for temperatures $T\geq 0.4$ due to a smaller range of $L$ and $N$ as compared to smaller temperatures, we restricted the ansatz \eqref{eq:extrapolation_ansatz} for the large $N$ and continuum extrapolations to be linear in $1/L$ and $1/N^2$ only. This should provide a good estimate for the energies, but suffers from a systematic error due to the missing higher order terms and an underestimation of the error bars.

\begin{figure}[ht]
	\centering 
	\includegraphics[scale=0.62]{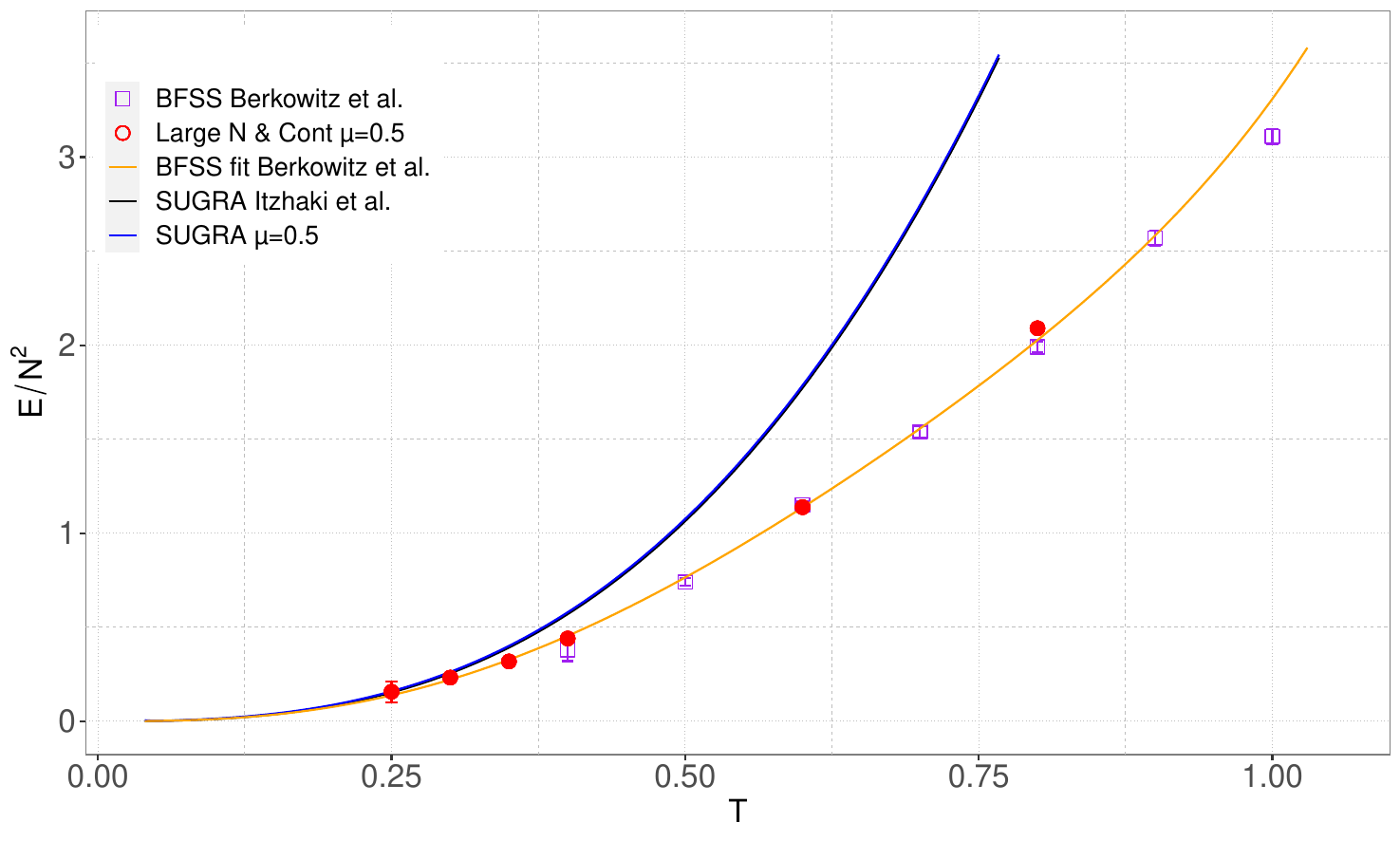}
	\caption{We are showing all the points we have simulated so far for $\mu=0.5$ along with the BFSS points from \cite{Berkowitz:2016jlq}. Even though the $\mu=0.5$ points for $T>0.3$ (see Table~\ref{tab:FitT>0.3largeNSfreecN}) have considerable systematic errors due to the simplified extrapolations, they are consistent with the BFSS points plus finite $\mu$ corrections.}
	\label{fig:EvsT_all_points}
\end{figure}

In Table~\ref{tab:FitT>0.3largeNSfreecN} we show the results of the large $N$ continuum fits. 
The results are plotted in Fig.~\ref{fig:EvsT_all_points} showing a good comparison with the BFSS points.  

\begin{table}[ht]
	\centering
	\begin{tabular}{| l || l | l | l || l |}
		\hline
		$T=0.35$ & \multicolumn{3}{c||}{19 ~ d.o.f} & 1.16 RSE \\ 
		\hline
		coefficient & \multicolumn{3}{c||}{fit} & Estimate based  \\ 
		& value & error & t-value  &  ~~on Ref.~ \cite{Berkowitz:2016jlq} \\
		\hline
		\hline
		$\varepsilon_{0,0}$     & 0.318  &   0.019 &  16.82 &  0.334  \\ \hline
		$\varepsilon_{1,0}$   & -1.38   &  1.15  & -1.20  &  -4.13 \\ \hline
		$\varepsilon_{0,1}$    & 10.66 &  1.62   & 6.58  & none \\ \hline 
		$\varepsilon_{0,2}$    & -71.34 &  33   & -2.10  & none \\
		\hline
		\hline
		$T=0.4$ & \multicolumn{3}{c||}{18 ~ d.o.f} & 1.05 RSE \\ 
		\hline
		coefficient & \multicolumn{3}{c||}{fit} & Estimate based  \\ 
		& value & error & t-value  &  ~~on Ref.~ \cite{Berkowitz:2016jlq} \\
		\hline
		\hline
		$\varepsilon_{0,0}$     & 0.44  &   0.02  &  21.92 &  0.460  \\ \hline
		$\varepsilon_{1,0}$   & 0.94   &  1.2  & 0.76  &  -4.47 \\ \hline
		$\varepsilon_{0,1}$    & 8.6 &  1.7   & 4.96  & none \\ \hline		
		$\varepsilon_{0,2}$    & -33.33 &  35   & -0.95  & none \\
		\hline
		$T=0.6$ & \multicolumn{3}{c||}{9 ~ d.o.f} & 1.62 RSE \\ 
		\hline
		coefficient & \multicolumn{3}{c||}{fit} & Estimate based  \\ 
		& value & error & t-value  &  ~~on Ref.~ \cite{Berkowitz:2016jlq} \\
		\hline
		\hline
		$\varepsilon_{0,0}$     & 1.138  &   0.017  &  65.81 &  1.132  \\ \hline
		$\varepsilon_{1,0}$   & -3.43   &  3.5  & -0.96  &  -5.84 \\ \hline
		$\varepsilon_{0,1}$    & 5.67 &  0.51   & 11.04  & none \\ \hline
		\hline
		$T=0.8$ & \multicolumn{3}{c||}{12 ~ d.o.f} & 1.12 RSE \\ 
		\hline
		coefficient & \multicolumn{3}{c||}{fit} & Estimate based  \\ 
		& value & error & t-value  &  ~~on Ref.~ \cite{Berkowitz:2016jlq} \\
		\hline
		\hline
		$\varepsilon_{0,0}$     & 2.09  &   0.02  &  105.11 &  1.992  \\ \hline
		$\varepsilon_{1,0}$   & 1.14   &  4.04  & 0.28  &  -7.42 \\ \hline
		$\varepsilon_{0,1}$    & 3.54 &  0.6   & 5.9  & none \\ \hline
		\hline
	\end{tabular}
	\caption{In this table we collect simultaneous large $N$ and continuum fits for $T>0.3$ temperatures at $\mu=0.5$. All other coefficients in \eqref{eq:extrapolation_ansatz} are set to zero. Residual standard errors (RSE) and degrees of freedom (d.o.f) are shown. As noted in the main text, the energies suffer from a systematic error due to the limited fitting ansatz.}
	\label{tab:FitT>0.3largeNSfreecN}
\end{table}

\subsection{Lower temperatures \texorpdfstring{($T <0.3$)}{(T<0.3)}} \label{sec:Tsmaller}
As reported in Ref.~\cite{Bergner:2021goh}, strong hysteresis is observed at temperatures around $T=0.25$ for the typical values of $\mu, N, L$ used in our simulations due to the existence of confined and deconfined phases. To study the deconfined phase, we must prevent tunneling to the confined phase. This may be achieved by a) increasing $N$, b) restarting simulations before a tunnelling event with a different set of random numbers, or c) by implementing constraints on the Polyakov loop~\cite{Bergner:2021goh}. Option a) is generally favoured as it is theoretically the cleanest. It is only partially feasible though as in general larger $N$ are numerically much harder. Option b) has the problem of requiring constant monitoring and frequent manual tempering with the simulations. Option c) does not have this problem, but is theoretically the least preferred because it is difficult to estimate the influence of the imposed constraints on the observables. 

We will first work with option a) in Sec.~\ref{sec:lowTunconstrained} for $N=16$ and then use option c) in Sec.~\ref{sec:lowTconstrained} for $N=18,21$. Thereby, we have a clean setup for $N=16$ and can test whether the constraints alter the simulation results significantly by performing a simultaneous large $N$ continuum extrapolation using all data and comparing it to gravity predictions. 

\subsubsection{Unconstrained simulation} \label{sec:lowTunconstrained}

\begin{figure}[ht!]
	\centering 
	\includegraphics[scale=0.43]{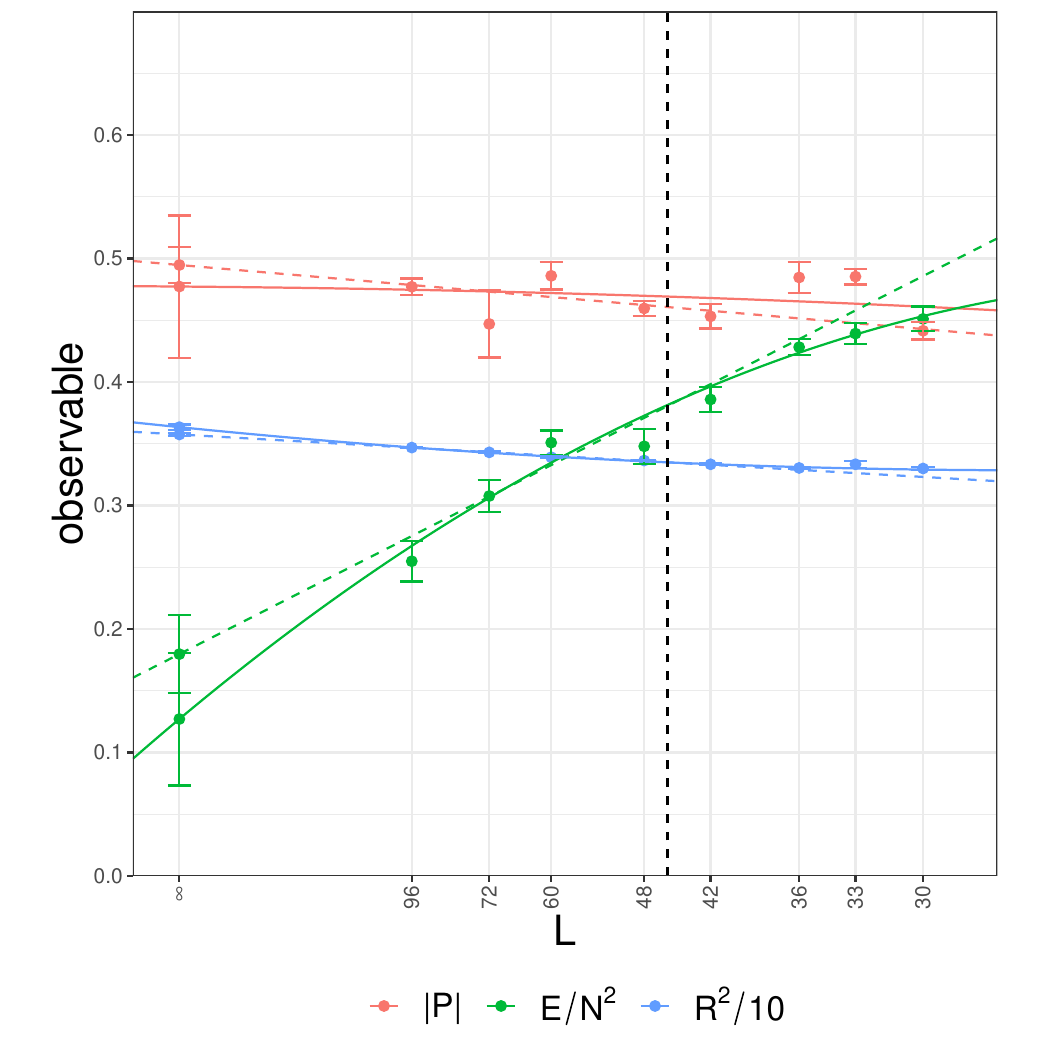}
	\includegraphics[scale=0.43]{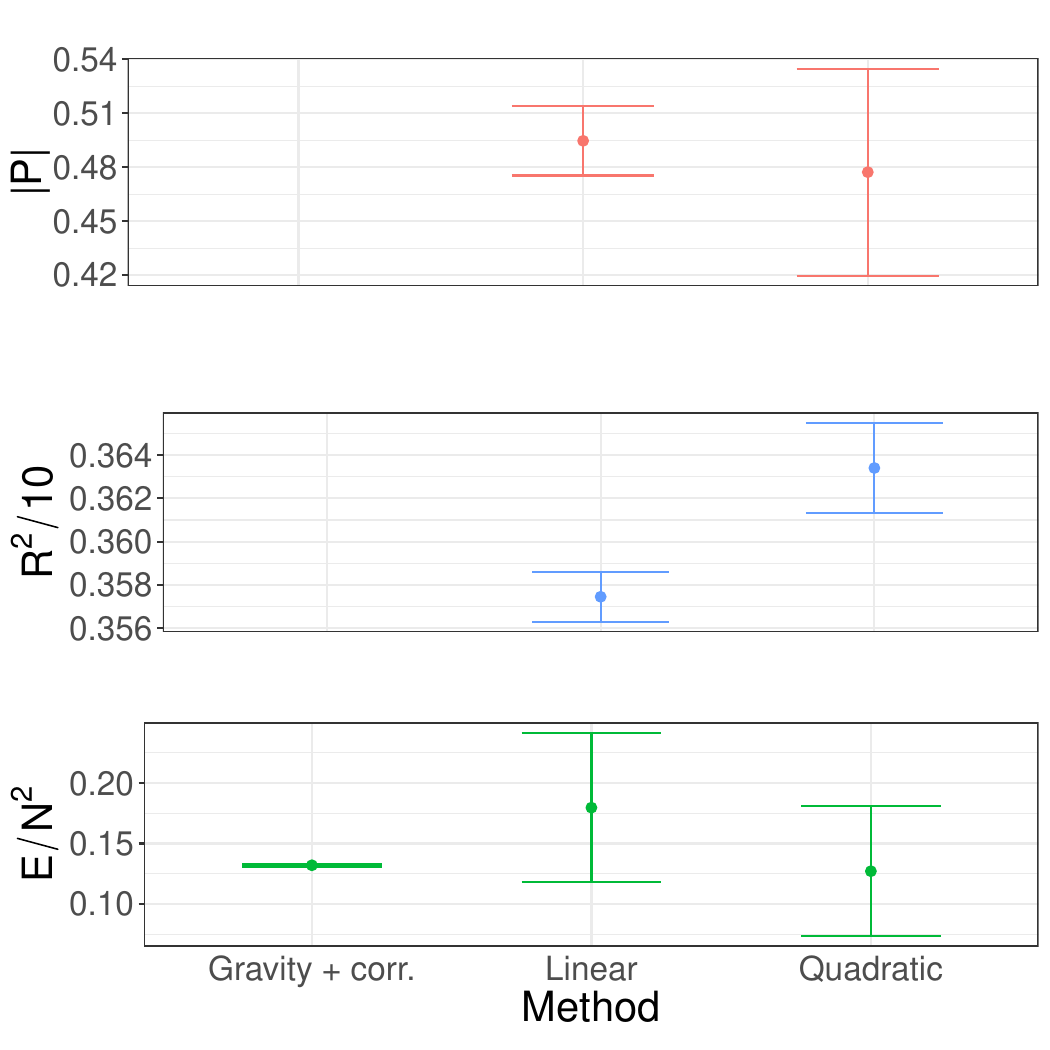}
	\caption{[Left] Large $L$ extrapolation for $T=0.25$, $N=16$, $\mu=0.5$ without constraints imposed. Solid lines are quadratic fits in $1/L$. Dashed lines are linear fits in $1/L$ taking into account data points left of the dashed black line. These extrapolation ans{\"a}tze lead to a consistent extrapolation value. [Right] Results of the continuum extrapolation, including the estimated gravity prediction of the energy. For the energy, we show the value obtained from the fit of the BFSS model at $T \geq 0.4$ based on \eqref{eq:BMN_Energy_expansion_large_N}, i.e. using the values for $a_1, a_2, b_1$ as before. As opposed to $T=0.3$, we don't include $1/N^4$ corrections since we were not able to estimate the coefficient $\varepsilon_{2,0}$ at $T=0.25$.}
	\label{fig:LargeSN16T0.25}
\end{figure}

We generally observe tunnelling at $N=16$, $T=0.25$ after a few thousand trajectories, which is enough to get a rough estimate of the energy. As initial configurations, we used either cold starts or forked the Monte Carlo chains from $T=0.3$. In both cases, we discarded sufficiently many configurations so that the correlation to the initial configuration is erased. For $N=16$, we found good agreement with the gravity prediction, although with sizeable error bars, see Fig.~\ref{fig:LargeSN16T0.25}. 

\subsubsection{Constrained simulation}  \label{sec:lowTconstrained}
The constraint simulation concerned the value of the Polyakov loop, which was fixed to 0.4 while allowing for a small fluctuation width. In this way, we are forcing the simulation to stay in the desired deconfined phase.

When including the data points at $N=18,21$ from the constrained simulations, it is possible to again to a simultaneous large-$N$ and continuum fit. The results are summarized in table \ref{tab:FitT0.25largeNSfreecN}, showing agreement with the estimated gravity prediction (finite $\mu$ and $\alpha'$-corrections from a numerical fit in \cite{Berkowitz:2016jlq}) within error bars. 
During the constrained simulations, the constraint term was in effect most of the time. Somewhat surprisingly, this does not seem to alter the expected result measurably. A similar observation was made in the bosonic case in \cite{Watanabe:2020ufk}.

\begin{table}[ht]
	\centering
	\begin{tabular}{| l || l | l | l || l |}
		\hline
		coefficient & \multicolumn{3}{c||}{fit} & Estimate based  \\ 
		& value & error & t-value  &  ~~on Ref.~ \cite{Berkowitz:2016jlq} \\
		\hline
		\hline
		$\varepsilon_{0,0}$     & 0.157  &   0.056  &  2.79 &  0.144  \\ \hline
		$\varepsilon_{1,0}$   & -5.29   &  10.9  & -0.49  &  -3.48 \\ \hline
		$\varepsilon_{0,1}$    &14.2  &   4.3   & 3.11  & none \\ \hline
		$\varepsilon_{0,2}$ &-138  &  93 &  -1.48  & none \\ \hline
	\end{tabular}
	\caption{Simultaneous large $N$ and continuum fit for $T=0.25$, $\mu=0.5$. All other coefficients in \eqref{eq:extrapolation_ansatz} are set to zero. Residual standard error: 1.13 on 8 degrees of freedom. Data includes unconstrained simulations at $N=16$ and constrained simulations at $N=18,21$. A Kolmogorov-Smirnov test as in Fig.~\ref{fig:LargeNSKS} is passed successfully, showing that the fitting ansatz describes the data well. We restricted to only first-order corrections in $1/N^2$ due to using $N\geq 16$ and omitted $\varepsilon_{1,1}$ as it was estimated to be very small at $T=0.3$.}
	\label{tab:FitT0.25largeNSfreecN}
\end{table}

\subsection{Comparison with simulations at \texorpdfstring{$\mu=0$}{μ=0}} \label{sec:ComparisonBFSS}

We were able to perform a limited set of simulations at $N \geq 32$ using an optimized GPU version of the simulation code. These values of $N$ turned out to be large enough to tame the flat directions at $T=0.3$ for $\mu=0$ so that the pure BFSS model could be directly simulated. Table \ref{tab:BFSSComparison} shows the simulation results along with a comparison to the BMN simulations at $\mu=0.5$ extrapolated to $\mu=0$ using the gravity prediction. We observe excellent agreement up to a single outlier, providing further evidence for the suitability of the approach taken in this paper.

\begin{table}[h]
	\centering
	\begin{tabular}{ | l | l | l | l | l |}
		\hline
		$N$ & $L$ & $E(\mu=0)$ & $\sigma_E$  &  relative error\\
		\hline
		\hline
		32    & 32  &   0.496  &  0.014 &  0.47  \\ \hline
		32    & 48  &   0.438  &  0.023 &  -0.89  \\ \hline
		48    & 24  &   0.560  &  0.021 &  0.83  \\ \hline
		48    & 32  &   0.467  &  0.035 &  1.10  \\ \hline
		48    & 48  &   0.397  &  0.025 &  0.98  \\ \hline
		64    & 32  &   0.504  &  0.020 &  0.09  \\ \hline
		32    & 64  &   0.364  &  0.033 &  0.32  \\ \hline
	\end{tabular}
	\caption{Simulation results for $T=0.3$, $\mu=0$ are summarized. We also present the relative error to the simulation results at $\mu=0.5$ as follows: we use the ansatz \eqref{eq:quadratic_ansatz} with the coefficients from table \ref{tab:FitT0.3largeNSfreecNN} to estimate the energy $E(\mu=0.5)$ at a given $N$, $L$ and correct for the finite $\mu$ correction using equation \eqref{eq:BMN_sugra_energy}, which amounts to subtracting $0.0076$ to get the $\mu=0$ value as opposed to $\mu=0.5$. We then subtract the BFSS measurement $E(\mu=0)$ and divide by its ($1\sigma$) statistical error $\sigma_E$. }
	\label{tab:BFSSComparison}
\end{table}

\subsection{\texorpdfstring{$E$}{E} vs \texorpdfstring{$T$}{T}} \label{sec:EvsT}

Based on our measurements, we are in a position to update the estimates for $a_1$ and $a_2$. For this, we use the data from \cite{Berkowitz:2016jlq} (with $0.4 \leq T<1.0$, see \cite{Berkowitz:2016jlq}) along with our data at $\mu=0.5$ for $T \leq 0.3$. We fix $a_0$ to its analytical value $7.41$ and estimate $a_1$ and $a_2$ based on the ansatz \eqref{eq:BMN_Energy_expansion_large_N}, including finite $\mu$ corrections for the $\mu=0.5$ data points. We obtain the updated values 
\begin{equation}
a_1 = -9.90 \pm 0.31, ~~~~ a_2 = 5.78 \pm 0.38 
\end{equation}
which are, as expected, consistent with \eqref{eq:estimates_a1a2} and with somewhat smaller error bars. 

The corresponding energy vs temperature plot was already shown in Sec.~\ref{sec:MainResult}. Since our values for $a_1$ and $a_2$ differ only marginally from those obtained in \cite{Berkowitz:2016jlq}, we refrain from replotting $E$ vs $T$.

\begin{figure}[ht]
	\centering 
	\includegraphics[scale=0.62]{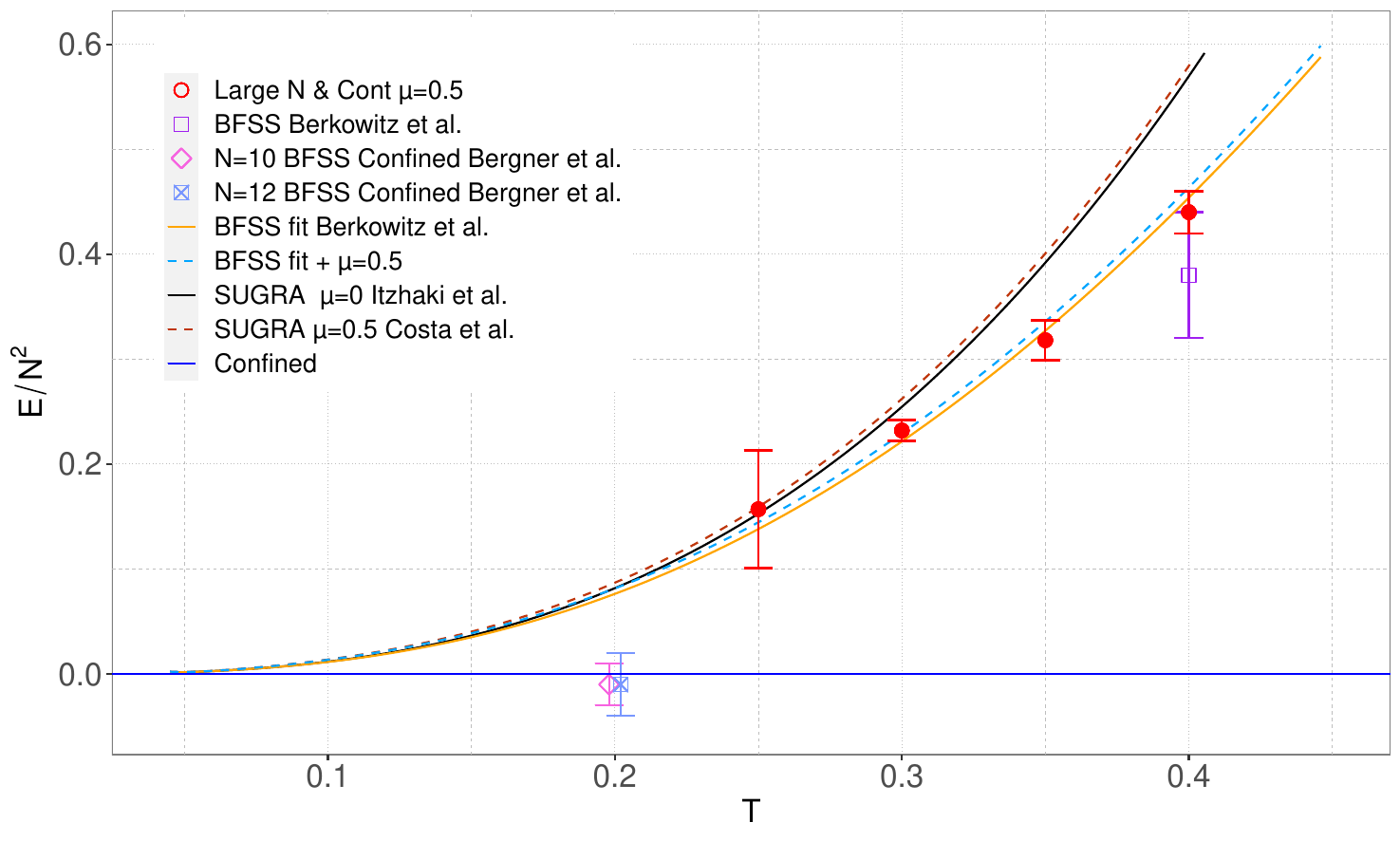}
	\caption{ Zoom-in view of the low-temperature region including the data point in the confined phase from Ref.~\cite{Bergner:2021goh}. The effect of the finite value of $\mu$ ($\mu=0.5$) is very small and practically negligible. The energy in the confined phase ($\frac{E}{N^2}\to 0$ as $N\to\infty$) is clearly different from that in the deconfined phase ($\frac{E}{N^2}\simeq 7.41 T^{14/5}$). 
	}\label{fig:BFSS_vs_BMN_small_T_con}
\end{figure}

In addition, we are presenting here also the confined phase for BFSS observed in Ref.~\cite{Bergner:2021goh} in Fig.~\ref{fig:BFSS_vs_BMN_small_T_con}. There is a clear difference between the energies of the confined and deconfined phases.  

\section{Conclusion and discussion}\label{sec:conclusion}

In this paper, we studied the low-temperature region of the duality between the D0-brane matrix model and its gravity dual.
To circumvent the difficulty associated with the flat directions, we used the BMN matrix model, which is a deformation of the BFSS matrix model by the flux parameter $\mu$. 
The stability of finite-$\mu$ simulations played an important role in the study of the low-temperature region that was not accessible in the past. 
We also showed both analytically and numerically that for finite yet small $\mu$ the difference between the energies in the BMN and BFSS models is small and indistinguishable within our simulation error bars. 

This is the first time the low-temperature region has been explored systematically. 
The $\alpha'$-correction to supergravity is 13\% or less at $T\le 0.3\lambda^{1/3}$. 
The simulation results are consistent with superstring theory in this temperature range and this gives us more confidence that the duality in the D0-matrix models is well under control in this region. 
We also managed to estimate the finite-$N$ corrections, in particular the term of order $N^{-4}$. 

This systematic study opens a new possibility to further put the gauge/gravity duality to a non-perturbative and numerical test in a region where both models can be studied. 
It will be an outstanding challenge in the future to probe even lower temperatures where $\alpha'$-corrections will be almost absent. 
At the same time, perturbative computations perhaps along the lines of Ref.~\cite{HyakutakeQuantumNear} could also be performed a priori and either verify or falsify the coefficient of the two-loop correction we are proposing here non-perturbatively. 

As shown in Fig.~\ref{fig:BFSS_vs_BMN_small_T_con}, the confined phase exists at low temperature~\cite{Bergner:2021goh}. 
To test the duality between the D0-brane matrix model and type IIA superstring theory, it is important to stay in the deconfined phase. 
At low temperatures, this requires us to use larger $N$ or a constraint on $E$ or $P$. 
The confined phase may describe M-theory~\cite{Bergner:2021goh} and understanding the gravity dual of the confined phase precisely is of great theoretical and conceptual interest.

\section*{Acknowledgments}

The authors would like to thank Joao Penedones, Jorge Santos and Toby Wiseman for discussions and comments.

G.~B. acknowledges support from the Deutsche Forschungsgemeinschaft (DFG) Grant No. BE 5942/3-1.
N.~B. and S.~P. were supported by an International Junior Research Group grant of the Elite Network of Bavaria. 
E.~R. is supported by Nippon Telegraph and Telephone Corporation (NTT) Research.
H.~W. is supported in part by the JSPS KAKENHI Grant Number JP 21J13014. 
M.~H. thanks the STFC Ernest Rutherford Grant ST/R003599/1.
A.~S. thanks the University of the Basque Country, Bilbao, for hospitality.
G.~B., S.~P. and M.~H. was partly supported in part by the International Centre for Theoretical Sciences (ICTS) for participating in the program ``Nonperturbative and Numerical Approaches to Quantum Gravity, String Theory and Holography" (code: ICTS/numstrings-2022/9). 
P.~V. acknowledges the support of the DOE under contract No.~DE-AC52-07NA27344 (Lawrence Livermore National Laboratory, LLNL).
The numerical simulations were performed on ATHENE, the HPC cluster of the Regensburg University Compute Centre and QPACE4, and on the Pascal supercomputer at LLNL.
We thank the LLNL Multiprogrammatic and Institutional Computing program for Grand Challenge super-computing allocations.

\section*{Data management}
No additional research data beyond the data presented and cited in this work are needed to validate the research findings in this work. Simulation data will be publicly available after publication.

\bibliographystyle{JHEP}
\bibliography{matrix-model}

\providecommand{\href}[2]{#2}\begingroup\raggedright\begin{thebibliography}{10}

\bibitem{Maldacena:1997re}
J.M.~Maldacena, \emph{{The Large N limit of superconformal field theories and
  supergravity}}, \href{https://doi.org/10.1023/A:1026654312961,
  10.4310/ATMP.1998.v2.n2.a1}{\emph{Int. J. Theor. Phys.} {\bfseries 38} (1999)
  1113} [\href{https://arxiv.org/abs/hep-th/9711200}{{\ttfamily
  hep-th/9711200}}].

\bibitem{Itzhaki:1998dd}
N.~Itzhaki, J.M.~Maldacena, J.~Sonnenschein and S.~Yankielowicz,
  \emph{{Supergravity and the large N limit of theories with sixteen
  supercharges}}, \href{https://doi.org/10.1103/PhysRevD.58.046004}{\emph{Phys.
  Rev.} {\bfseries D58} (1998) 046004}
  [\href{https://arxiv.org/abs/hep-th/9802042}{{\ttfamily hep-th/9802042}}].

\bibitem{Banks:1996vh}
T.~Banks, W.~Fischler, S.H.~Shenker and L.~Susskind, \emph{{M theory as a
  matrix model: A Conjecture}},
  \href{https://doi.org/10.1103/PhysRevD.55.5112}{\emph{Phys. Rev.} {\bfseries
  D55} (1997) 5112} [\href{https://arxiv.org/abs/hep-th/9610043}{{\ttfamily
  hep-th/9610043}}].

\bibitem{deWit:1988wri}
B.~de~Wit, J.~Hoppe and H.~Nicolai, \emph{{On the Quantum Mechanics of
  Supermembranes}},
  \href{https://doi.org/10.1016/0550-3213(88)90116-2}{\emph{Nucl. Phys.}
  {\bfseries B305} (1988) 545}.

\bibitem{Bergner:2021goh}
{\scshape MCSMC} collaboration, \emph{{Confinement/deconfinement transition in
  the D0-brane matrix model -- A signature of M-theory?}},
  \href{https://doi.org/10.1007/JHEP05(2022)096}{\emph{JHEP} {\bfseries 05}
  (2022) 096} [\href{https://arxiv.org/abs/2110.01312}{{\ttfamily
  2110.01312}}].

\bibitem{Anagnostopoulos:2007fw}
K.N.~Anagnostopoulos, M.~Hanada, J.~Nishimura and S.~Takeuchi, \emph{{Monte
  Carlo studies of supersymmetric matrix quantum mechanics with sixteen
  supercharges at finite temperature}},
  \href{https://doi.org/10.1103/PhysRevLett.100.021601}{\emph{Phys. Rev. Lett.}
  {\bfseries 100} (2008) 021601}
  [\href{https://arxiv.org/abs/0707.4454}{{\ttfamily 0707.4454}}].

\bibitem{Catterall:2008yz}
S.~Catterall and T.~Wiseman, \emph{{Black hole thermodynamics from simulations
  of lattice Yang-Mills theory}},
  \href{https://doi.org/10.1103/PhysRevD.78.041502}{\emph{Phys. Rev.}
  {\bfseries D78} (2008) 041502}
  [\href{https://arxiv.org/abs/0803.4273}{{\ttfamily 0803.4273}}].

\bibitem{Hanada:2008ez}
M.~Hanada, Y.~Hyakutake, J.~Nishimura and S.~Takeuchi, \emph{{Higher derivative
  corrections to black hole thermodynamics from supersymmetric matrix quantum
  mechanics}},
  \href{https://doi.org/10.1103/PhysRevLett.102.191602}{\emph{Phys. Rev. Lett.}
  {\bfseries 102} (2009) 191602}
  [\href{https://arxiv.org/abs/0811.3102}{{\ttfamily 0811.3102}}].

\bibitem{Hanada:2008gy}
M.~Hanada, A.~Miwa, J.~Nishimura and S.~Takeuchi, \emph{{Schwarzschild radius
  from Monte Carlo calculation of the Wilson loop in supersymmetric matrix
  quantum mechanics}},
  \href{https://doi.org/10.1103/PhysRevLett.102.181602}{\emph{Phys. Rev. Lett.}
  {\bfseries 102} (2009) 181602}
  [\href{https://arxiv.org/abs/0811.2081}{{\ttfamily 0811.2081}}].

\bibitem{Hanada:2013rga}
M.~Hanada, Y.~Hyakutake, G.~Ishiki and J.~Nishimura, \emph{{Holographic
  description of quantum black hole on a computer}},
  \href{https://doi.org/10.1126/science.1250122}{\emph{Science} {\bfseries 344}
  (2014) 882} [\href{https://arxiv.org/abs/1311.5607}{{\ttfamily 1311.5607}}].

\bibitem{Kadoh:2015mka}
D.~Kadoh and S.~Kamata, \emph{{Gauge/gravity duality and lattice simulations of
  one dimensional SYM with sixteen supercharges}},
  \href{https://arxiv.org/abs/1503.08499}{{\ttfamily 1503.08499}}.

\bibitem{Filev:2015hia}
V.G.~Filev and D.~O'Connor, \emph{{The BFSS model on the lattice}},
  \href{https://doi.org/10.1007/JHEP05(2016)167}{\emph{JHEP} {\bfseries 05}
  (2016) 167} [\href{https://arxiv.org/abs/1506.01366}{{\ttfamily
  1506.01366}}].

\bibitem{Berkowitz:2016jlq}
E.~Berkowitz, E.~Rinaldi, M.~Hanada, G.~Ishiki, S.~Shimasaki and P.~Vranas,
  \emph{{Precision lattice test of the gauge/gravity duality at large-$N$}},
  \href{https://doi.org/10.1103/PhysRevD.94.094501}{\emph{Phys. Rev.}
  {\bfseries D94} (2016) 094501}
  [\href{https://arxiv.org/abs/1606.04951}{{\ttfamily 1606.04951}}].

\bibitem{Rinaldi:2017mjl}
E.~Rinaldi, E.~Berkowitz, M.~Hanada, J.~Maltz and P.~Vranas, \emph{{Toward
  Holographic Reconstruction of Bulk Geometry from Lattice Simulations}},
  \href{https://doi.org/10.1007/JHEP02(2018)042}{\emph{JHEP} {\bfseries 02}
  (2018) 042} [\href{https://arxiv.org/abs/1709.01932}{{\ttfamily
  1709.01932}}].

\bibitem{Catterall:2010fx}
S.~Catterall, A.~Joseph and T.~Wiseman, \emph{{Thermal phases of D1-branes on a
  circle from lattice super Yang-Mills}},
  \href{https://doi.org/10.1007/JHEP12(2010)022}{\emph{JHEP} {\bfseries 12}
  (2010) 022} [\href{https://arxiv.org/abs/1008.4964}{{\ttfamily 1008.4964}}].

\bibitem{Catterall:2017lub}
S.~Catterall, R.G.~Jha, D.~Schaich and T.~Wiseman, \emph{{Testing holography
  using lattice super-Yang-Mills theory on a 2-torus}},
  \href{https://doi.org/10.1103/PhysRevD.97.086020}{\emph{Phys. Rev.}
  {\bfseries D97} (2018) 086020}
  [\href{https://arxiv.org/abs/1709.07025}{{\ttfamily 1709.07025}}].

\bibitem{Catterall:2020nmn}
S.~Catterall, J.~Giedt, R.G.~Jha, D.~Schaich and T.~Wiseman,
  \emph{{Three-dimensional super-Yang--Mills theory on the lattice and dual
  black branes}},  \href{https://arxiv.org/abs/2010.00026}{{\ttfamily
  2010.00026}}.

\bibitem{Berenstein:2002jq}
D.E.~Berenstein, J.M.~Maldacena and H.S.~Nastase, \emph{{Strings in flat space
  and pp waves from N=4 superYang-Mills}},
  \href{https://doi.org/10.1088/1126-6708/2002/04/013}{\emph{JHEP} {\bfseries
  04} (2002) 013} [\href{https://arxiv.org/abs/hep-th/0202021}{{\ttfamily
  hep-th/0202021}}].

\bibitem{Costa:2014wya}
M.S.~Costa, L.~Greenspan, J.~Penedones and J.~Santos, \emph{{Thermodynamics of
  the BMN matrix model at strong coupling}},
  \href{https://doi.org/10.1007/JHEP03(2015)069}{\emph{JHEP} {\bfseries 03}
  (2015) 069} [\href{https://arxiv.org/abs/1411.5541}{{\ttfamily 1411.5541}}].

\bibitem{Dhindsa:2022vch}
N.S.~Dhindsa, A.~Joseph, A.~Samlodia and D.~Schaich, \emph{{Non-perturbative
  phase structure of the bosonic BMN matrix model}},
  \href{https://arxiv.org/abs/2201.08791}{{\ttfamily 2201.08791}}.

\bibitem{Schaich:2022duk}
D.~Schaich, R.G.~Jha and A.~Joseph, \emph{{Thermal phase structure of
  dimensionally reduced super-Yang-Mills}},
  \href{https://arxiv.org/abs/2201.03097}{{\ttfamily 2201.03097}}.

\bibitem{Pateloudis:2022GvsU}
S.~Pateloudis, G.~Bergner, N.~Bodendorfer, M.~Hanada, E.~Rinaldi and
  A.~Sch\"afer, \emph{{Nonperturbative test of the Maldacena-Milekhin
  conjecture for the BMN matrix model}},
  \href{https://doi.org/10.1007/JHEP08(2022)178}{\emph{JHEP} {\bfseries 08}
  (2022) 178} [\href{https://arxiv.org/abs/2205.06098}{{\ttfamily
  2205.06098}}].

\bibitem{HyakutakeQuantumNear}
Y.~{Hyakutake}, \emph{{Quantum near-horizon geometry of a black 0-brane}},
  \href{https://doi.org/10.1093/ptep/ptu028}{\emph{Progress of Theoretical and
  Experimental Physics} {\bfseries 2014} (2014) 033B04}
  [\href{https://arxiv.org/abs/1311.7526}{{\ttfamily 1311.7526}}].

\bibitem{Hanada:2007ti}
M.~Hanada, J.~Nishimura and S.~Takeuchi, \emph{{Non-lattice simulation for
  supersymmetric gauge theories in one dimension}},
  \href{https://doi.org/10.1103/PhysRevLett.99.161602}{\emph{Phys. Rev. Lett.}
  {\bfseries 99} (2007) 161602}
  [\href{https://arxiv.org/abs/0706.1647}{{\ttfamily 0706.1647}}].

\bibitem{Catterall:2007fp}
S.~Catterall and T.~Wiseman, \emph{{Towards lattice simulation of the gauge
  theory duals to black holes and hot strings}},
  \href{https://doi.org/10.1088/1126-6708/2007/12/104}{\emph{JHEP} {\bfseries
  12} (2007) 104} [\href{https://arxiv.org/abs/0706.3518}{{\ttfamily
  0706.3518}}].

\bibitem{Hanada:2021ipb}
M.~Hanada, \emph{{Bulk geometry in gauge/gravity duality and color degrees of
  freedom}}, \href{https://doi.org/10.1103/PhysRevD.103.106007}{\emph{Phys.
  Rev. D} {\bfseries 103} (2021) 106007}
  [\href{https://arxiv.org/abs/2102.08982}{{\ttfamily 2102.08982}}].

\bibitem{KlebanovEntropyOfNear}
I.~Klebanov and A.~Tseytlin, \emph{Entropy of near-extremal black p-branes},
  \href{https://doi.org/https://doi.org/10.1016/0550-3213(96)00295-7}{\emph{Nuclear
  Physics B} {\bfseries 475} (1996) 164}.

\bibitem{HyakutakeQuantumMwave}
Y.~{Hyakutake}, \emph{{Quantum M-wave and black 0-brane}},
  \href{https://doi.org/10.1007/JHEP09(2014)075}{\emph{Journal of High Energy
  Physics} {\bfseries 2014} (2014) 75}
  [\href{https://arxiv.org/abs/1407.6023}{{\ttfamily 1407.6023}}].

\bibitem{Catterall:2009xn}
S.~Catterall and T.~Wiseman, \emph{{Extracting black hole physics from the
  lattice}}, \href{https://doi.org/10.1007/JHEP04(2010)077}{\emph{JHEP}
  {\bfseries 04} (2010) 077} [\href{https://arxiv.org/abs/0909.4947}{{\ttfamily
  0909.4947}}].

\bibitem{Berkowitz:2016znt}
E.~Berkowitz, M.~Hanada and J.~Maltz, \emph{{Chaos in Matrix Models and Black
  Hole Evaporation}},
  \href{https://doi.org/10.1103/PhysRevD.94.126009}{\emph{Phys. Rev. D}
  {\bfseries 94} (2016) 126009}
  [\href{https://arxiv.org/abs/1602.01473}{{\ttfamily 1602.01473}}].

\bibitem{Watanabe:2020ufk}
H.~Watanabe, G.~Bergner, N.~Bodendorfer, S.~Shiba~Funai, M.~Hanada, E.~Rinaldi
  et~al., \emph{{Partial deconfinement at strong coupling on the lattice}},
  \href{https://doi.org/10.1007/JHEP02(2021)004}{\emph{JHEP} {\bfseries 02}
  (2021) 004} [\href{https://arxiv.org/abs/2005.04103}{{\ttfamily
  2005.04103}}].

\end{thebibliography}\endgroup

\end{document}